\documentclass[letterpaper,review,anonymous,twocolumn,10pt]{article}
\usepackage{usenix2019_v3}

\AtBeginDocument{%
  \providecommand\BibTeX{{%
    \normalfont B\kern-0.5em{\scshape i\kern-0.25em b}\kern-0.8em\TeX}}}

\usepackage{booktabs} 
\usepackage{filecontents}
\usepackage{subcaption}
\usepackage{xcolor}
\usepackage{comment}

\usepackage[many]{tcolorbox}
\usepackage{tikz}
\usepackage{url}
\usepackage{color}
\usepackage{bm}
\usepackage{enumitem}
\usepackage{diagbox}
\usepackage{listings}
\usepackage{inconsolata}
\definecolor{lightgray}{gray}{0.75}

\newcommand*\circled[1]{\tikz[baseline=(char.base)]{\node[shape=circle,fill=black,text=white,draw,inner sep=1pt,scale=0.8] (char) {#1};}}

\newcommand{\ttimes}{\! \times \!}

\newcommand{\fahad}[1]{\textcolor{blue}{Fahad: #1}}
\newcommand{\abd}[1]{\textcolor{red}{Abd: #1}}
\newcommand{\mohsin}[1]{\textcolor{red}{Mohsin: #1}}

\newcommand{\dlt}{DLT}
\newcommand{\sys}{\textsf{PCS}}
\newcommand{\cte}{\textsf{JCTpred}}

\newcommand{\ecte}{Pred$_{err}$}
\newcommand{\policy}{\sys{}}

\renewcommand{\paragraph}[1]{\medskip{}\noindent{}\textbf{#1}}
 
\begin{document}

\author{
{\rm Abdullah Bin Faisal}\\
Tufts University\\
\and
{\rm Noah Martin}\\
Tufts University\\
\and
{\rm Hafiz Mohsin Bashir}\\
Tufts University\\
\and
{\rm Swaminathan Lamelas}\\
Tufts University\\
\and
{\rm Fahad R. Dogar}\\
Tufts University\\
} 

\title{Towards providing reliable job completion time predictions using PCS}

\maketitle

\begin{abstract}
In this paper we build a case for providing job completion time predictions to cloud users, similar to the delivery date of a package or arrival time of a booked ride.
Our analysis reveals that providing predictability can come at the expense of performance and fairness.
Existing cloud scheduling systems optimize for extreme points in the trade-off space, making them either extremely unpredictable or impractical.

To address this challenge, we present \sys{}, a new scheduling framework that aims to provide predictability while balancing other traditional objectives.
The key idea behind \sys{} is to use Weighted-Fair-Queueing (WFQ) and find a suitable configuration of different WFQ parameters (e.g., class weights) that meets specific goals for predictability.
It uses a simulation-aided search strategy, to efficiently discover WFQ configurations that lie on the Pareto front of the trade-off space between these objectives.
We implement and evaluate \sys{} in the context of DNN job scheduling on GPUs.
Our evaluation, on a small scale GPU testbed and larger-scale simulations, shows that \sys{} can provide accurate completion time estimates while marginally compromising on performance and fairness.

\end{abstract}


\lstset{%
language=C++,
basicstyle=\medium
}

\section{Introduction}
\label{sec:intro}
Humans desire predictability in their daily lives~\cite{predictability-human-psyc}: from knowing how long their home-to-office commute will be to the arrival time of an Amazon package~\cite{onlineretail} or an Uber ride~\cite{uber}.
Fortunately, most real world systems (e.g., transportation, e-commerce etc) reflect this need by providing their users with a (reliable) prediction (e.g., estimated delivery date).
As more and more of our lives move to the cloud (e.g., Metaverse~\cite{metaverse,meta-aas}), it begs the question of whether the cloud can offer similar predictability.
More concretely, when a user submits a ``job'' (e.g., train a machine learning model) to the cloud, can the cloud provide a reliable job completion time prediction?


Such feedback can notably ensure a seamless experience and ease user frustration; perhaps more emphatically than simply making the cloud faster or fairer, according to studies in human psychology~\cite{how-tolerable-is-delay, delay-information} and systems usage~\cite{morpheus}.
It can also empower users to decide between different cloud platforms and services within a cloud based on the provided estimate, or be integrated with emerging inter-cloud brokers (e.g., SkyPilot~\cite{sky-computing}).
In light of this, we advocate for the need to provide reliable job completion time predictions as a \emph{core} primitive in today's cloud, making it more akin to real world systems we interact with.


Supporting this primitive depends on multiple aspects of the user-cloud ecosystem to be predictable (e.g., knowledge of job sizes).
The technical focus of this paper, assuming all else is predictable, is on understanding the (un)predictability of existing scheduling mechanisms used by cloud (sub)systems and we situate our study in the context of multi-tenant clusters (e.g., PAI~\cite{PAI}, Philly~\cite{philly}, etc) --- this is our definition of the cloud.

Multi-tenant clusters are an important and challenging scenario to focus on.
They are important because they enable academics or research and development teams to run their workloads in a feasible and cost-effective manner.
They also present a challenging scenario to deal with because unlike the public cloud setting where users can pay for dedicated (and hence predictable) resources or Service-Level-Agreements (SLAs), these clusters are best-effort.
They rely heavily on scheduling to determine how the underlying resources are to be shared amongst tasks/jobs, applications, or tenants (users)~\cite{mesos, omega, bistro, heracles, rans-hotnets, das}.
A large body of work has proposed various scheduling techniques that optimize for metrics like minimizing average/tail job completion times (JCT) (e.g., Tiresias~\cite{tiresias}), fairness (e.g., Themis~\cite{themis}), and meeting deadlines (e.g., Chronus~\cite{chronus}), for different cluster resources (e.g., GPU/CPU~\cite{themis, tiresias,AFS, shockwave}, network bandwidth~\cite{aalo, 2D, baraat, pias, pase}, etc).
While these metrics are useful, aggressively optimizing for them introduces non-determinism in the scheduling decision, making it challenging to provide reliable job completion time predictions.

Our key observation is that a scheduling policy's use of \emph{unbounded preemption} determines its (lack of) ability to provide reliable Job Completion Time Predictions (\cte{}).
Preemption is collectively defined as when some or all of the resources (e.g., GPUs) assigned to a job are taken away and assigned to another job or a queued job's expected execution order is changed due to a future arrival.
Preemption is a key enabler for scheduling policies based on size and deadlines~\cite{tiresias, chronus, optimus}, fairness and resource efficiency~\cite{themis, gandiva, gandiva_fair, AFS}, and handling dynamic jobs~\cite{pollux, shockwave, hypersched}.
While preemption is central to achieving the desired objectives of these schemes, it leads to unpredictability (prediction error) due to (repeated) preemptions from future jobs.
On the other hand, deterministic scheduling policies that lack preemption, such as First-In-First-Out (FIFO), are predictable (as future arrivals do not impact current jobs) but can result in extremely poor performance and a lack of fairness due to Head-Of-Line (HOL) blocking~\cite{baraat, lps, tiresias, 2D}.


This observation highlights that there is an inherent trade-off between offering predictability and optimizing for other metrics (e.g., minimizing JCTs).
Existing schedulers lie on extreme points of this trade-off space, making them either extremely unpredictable due to their use of unbounded preemption or impractical (no preemption).

This prompts us to ask: are there intermediate points on this trade-off space that can offer predictability while being practical; achieved using (varying degree of) \emph{bounded preemption}?
As we show later, it is possible to reduce the prediction error from 80\% to between 10-3\% while only making a marginal compromise (2-2.7$\ttimes{}$) on performance --- for reference, FIFO has a prediction error of 0\% at a 90$\ttimes{}$ performance cost.
Furthermore, the cloud system may want to operate at potentially any one of these trade-off points based on their \emph{relative} preferences between predictability and other objectives.
The trade-off space could be very large, and not all points would be useful.
How do we enable cloud operators to express their preferences and have a mechanism to discover the trade-off space in an efficient manner?

To address these questions, we propose a novel scheduling framework called Predictability-Centric Scheduling (\sys{}) that aims to provide reliable \cte{} (predictability) while balancing other practical goals (flexibility) such as performance and fairness.
\sys{} exposes a high level bi-directional preference interface which allows cloud operators to express the objectives they are interested in (e.g., avg JCTs vs avg prediction error).
To facilitate cloud operators in making an informed choice based on their \emph{relative} preferences, \sys{} provides a \emph{set} of Pareto-optimal trade-offs.
Each Pareto-optimal trade-off improves one objective (e.g., predictability) while marginally sacrificing on other objectives (e.g., performance and/or fairness).
This is unlike other tunable schedulers~\cite{gavel, decima, selftune} which typically return a single solution.


At its core, \sys{} leverages Weighted-Fair-Queuing (WFQ) as a basic building block~\cite{wfq-scott}.
Our use of WFQ is motivated by the fact that it uses bounded preemption and offers direct control over the extent of preemption used.
WFQ maps incoming jobs to a fixed number of classes, uses FIFO to schedule jobs within a class and assigns a guaranteed resource share (class weights) to each class.
These properties bound the preemptions and reordering experienced by jobs.
Furthermore, the number of classes and their assigned weights are tunable parameters of the WFQ policy.
This allows direct control over i) predictability (e.g., by creating limited number of classes), ii) performance (e.g., by assigning a higher weight to classes with smaller jobs), and iii) fairness (e.g., by assigning equal weights), motivating its flexibility and ability to achieve Pareto-optimal trade-offs.


Finding Pareto-optimal WFQ configurations is challenging because the space of possible configurations is large, each configuration potentially lying on one point in the trade-off space, with some trade-offs not feasible (e.g., optimal performance and maximum predictability) or beneficial (e.g., more unpredictable and unfairer than existing schemes).
To address this challenge, \sys{} uses a highly-parallel simulation-based search strategy with an intelligent parameterization of WFQ using heuristics, to efficiently find suitable and feasible (Pareto-optimal) WFQ configurations.
For example, we use the variation in job-sizes to determine the number of classes and thresholds as opposed to trying out arbitrary combinations.
We show that Pareto-optimal trade-offs can be discovered for realistic workloads in a timely manner (\S\ref{subsec:micro}).

A key benefit of \sys{} is that it is a generic scheduler, which can accommodate various types of jobs (e.g., network flows, DNN training jobs), allowing it to be realized in various multi-tenant scheduling scenarios.
It only requires a mapping between the execution time of a job and the resources allocated to it i.e., a job's demand function.
The demand function can either be provided by the user or estimated by the system~\cite{morpheus,themis,cilantro}.
\sys{} uses these demand functions to balance considerations for performance and fairness (e.g., when dealing with sub-linear scaling jobs) to be competitive with schedulers (e.g., AFS~\cite{AFS}) that aggressively use a job's demand function to optimize their scheduling decision, as we show in~\S\ref{sec:eval}. 

While \sys{} is potentially useful for any resource (e.g., network, storage, etc), it is particularly important for GPU scheduling for ML training workloads, an evaluation scenario we consider in this paper.
ML applications are becoming very important (e.g., LLMS~\cite{gpt}) and training them can take a significant time (e.g., hours to days) with high variability, and hence getting a reliable \cte{} becomes important.
We implement and evaluate \sys{} for realistic ML training workloads on a small-scale cluster as well as large scale simulations.
Our evaluation shows that \sys{} can successfully discover Pareto-optimal WFQ configurations to meet varying trade-offs.
For example, \sys{} can reduce the prediction error by 50-800\% while being within 1.1-3.5$\ttimes{}$ of performance and fairness optimal schemes (\S\ref{sec:eval}).

Overall, we make the following contributions:
\begin{itemize}[leftmargin=*]

    \item We show that state-of-the-art scheduling policies which optimize for performance and fairness~\cite{AFS, themis, tiresias, gandiva} result in unpredictability.
    Our analysis shows that these policies typically lie on extreme points of predictability-performance or predictability-fairness trade-offs~(\S\ref{sec:case}).
    
    \item We design \sys{}, a generic job scheduler, which uses WFQ in a unique and novel way to achieve predictability and flexibility~(\S\ref{subsec:policy}). 
        
    \item We provide a simple but expressive bi-directional interface to be used by cloud operators, enabling them to specify different high level objectives and giving them the ability to choose between trade-offs --- a property existing scheduling systems fail to provide~(\S\ref{subsec:preference_interface}).
    
    \item We implement a prototype of \sys{} in Ray and evaluate it on a testbed and in simulations for realistic DNN workloads~(\S\ref{sec:system} \S\ref{sec:eval}). 
    
\end{itemize}

\sys{} is a step in providing predictability in today's cloud systems.
It opens up important questions which we discuss in~\S\ref{sec:discussion}.
Finally, we build upon and benefit from a large body of prior work in scheduling systems, which we discuss in \S\ref{sec:related-work}.
The code for \sys{} is made available at \href{https://github.com/abdullahfsm/PCS}{https://github.com/abdullahfsm/PCS}.

\section{A Case for Predictable Scheduling} \label{sec:case}
In this section, we provide several use-cases of predictable scheduling, motivating the need for it to be a core primitive in today's cloud and show how it is different from deadline based systems.
Throughout this section, we try to draw analogies between real world systems and the cloud.

\paragraph{Why provide JCT predictions (\cte{})?}
A scheduling system that provides \cte{} can have two broad benefits:
\underline{(1) Alleviating User frustration.}
Several studies on real-world systems (e.g., online retail~\cite{online-shopping-usability}, airlines~\cite{airline-delays}) show that providing a timeline to users can help ease frustration in face of long and variable waiting times~\cite{delay-announcements, delay-information, delay-announcements-survey}.
\cte{} can offer a similar role in the context of cloud systems, where users can suffer from large and unpredictable slowdowns, inevitably leading to a poor and frustrating experience~\cite{how-tolerable-is-delay,frustration_still_common}.
Take into consideration the evidence gathered from several GPU cluster operators and users:
measurements show up to 100 hours of queuing and preemption related delays for ML training jobs at Microsoft~\cite{philly}.
A recent survey reveals that users are often trying to guess when their training jobs will complete and that user-driven predictions can be off by more than $100\%$ i.e. the system takes twice as long to complete their job compared to the user's expectation, with some users finding it \emph{impossible} to make any meaningful predictions~\cite{chronus}.
With the paradigm of AutoML, jobs that spawn hundreds of DNN trials~\cite{themis, hyperband}, and Large Language Models (e.g., GPT~\cite{gpt}) consisting of $O(trillion)$ parameters, becoming mainstream, these issues will only exacerbate~\cite{high-cost-of-AI}.


\underline{(2) Enabling decision making.}
In real-world systems, if the predicted timeline is long, customers may elect to perform other tasks or seek alternatives~\cite{waittimepsychology}.
For example, estimated delivery dates can help shoppers decide between e-commerce platforms (e.g., Amazon~\cite{amazon} vs Temu~\cite{temu}) and even between sellers within a platform.
Today's cloud users have similar choices to make and \cte{} can enable them to make these choices in a more informed way.
For example, it can help users decide between different cloud systems to run their jobs/applications on; each potentially offering a different cost-\cte{} trade-off.
As a forward looking avenue, \cte{} can facilitate the growing eco-system around inter-cloud brokers which orchestrate seamless access to multiple clouds with low user effort (e.g., SkyPilot~\cite{sky-computing, sky-computing2, sky-computing3, sky-computing4}).
Within a cloud, \cte{} can facilitate user applications in i) replica selection strategies (e.g., MittOS~\cite{mittos}), ii) selecting which model variant/pipeline to use for inference, based on the accuracy-\cte{} trade-off~\cite{supernet-inference,inferline,model-switch,unfoldml} in AI serving systems, and iii) optimizing the right parallelism and placement for network-bound data processing tasks~\cite{NEAT, jockey}.

\paragraph{Why deadlines are not the answer?}
One may wonder how the predictability metric is different from deadlines (and the large body of work on deadline-based scheduling~\cite{chronus,genie,cheng2015resource,li2015dcloud}) where a user accompanies a deadline along with their job and the system tries to satisfy it.
The fundamental difference is that in the deadline-based context the burden lies on the \emph{user} to provide a timeline to the system, with the system deciding the user's fate.
We posit that it should instead be the \emph{system} that provides the user with a timeline (i.e., a \cte{}), empowering them to decide whether it is acceptable or not.
Our approach is analogous to real-world systems like ride-sharing where most users request a ride, wanting it ASAP (i.e., no deadline) while the system comes up with the expected arrival time of the ride.

Even if we try to shoehorn predictability into deadlines, it will be challenging for two reasons. First, coming up with a reasonable deadline is hard because the slowdown of a job is highly dependent on: i) cluster load (which can be highly variable and bursty at short timescales) and ii) underlying  job-to-resource mapping which is (dynamically) determined at run time~\cite{morpheus} and can result in significant variation due to heterogeneity in the underlying resources (e.g., low vs. high end GPUs~\cite{PAI,gandiva_fair,gavel}, RDMA vs. TCP~\cite{habitat, bytescheduler}, etc.,). Second, unless there is an inherent difference in user requirements (and hence deadlines), users have the incentive to specify a small deadline (i.e., to act greedy), which limits any prioritization the system can enable. In both the above cases, the lack of reasonable deadlines will render the system ineffective.
\begin{figure*}[!t]
    \centering    
    \subfloat[Impact of future arrivals on scheduling]{\includegraphics[width=0.8\textwidth]{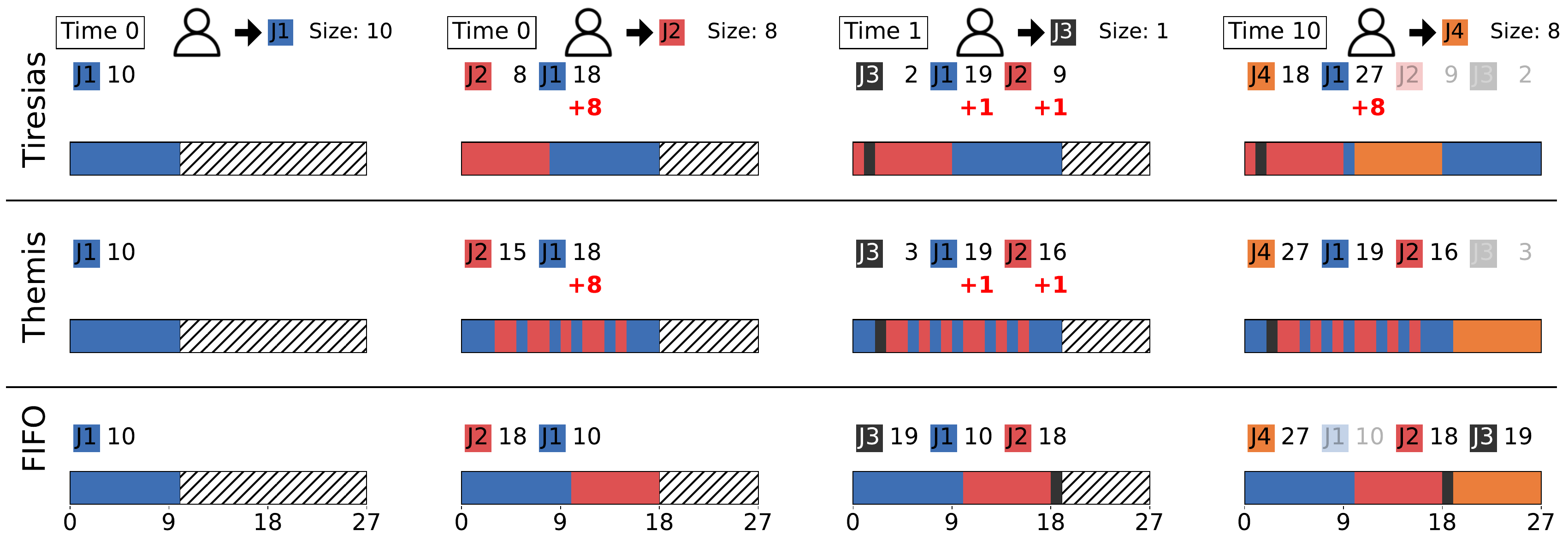}\label{subfig:toy-tiresias}}
    \subfloat[Summary]{\includegraphics[width=0.2\textwidth]{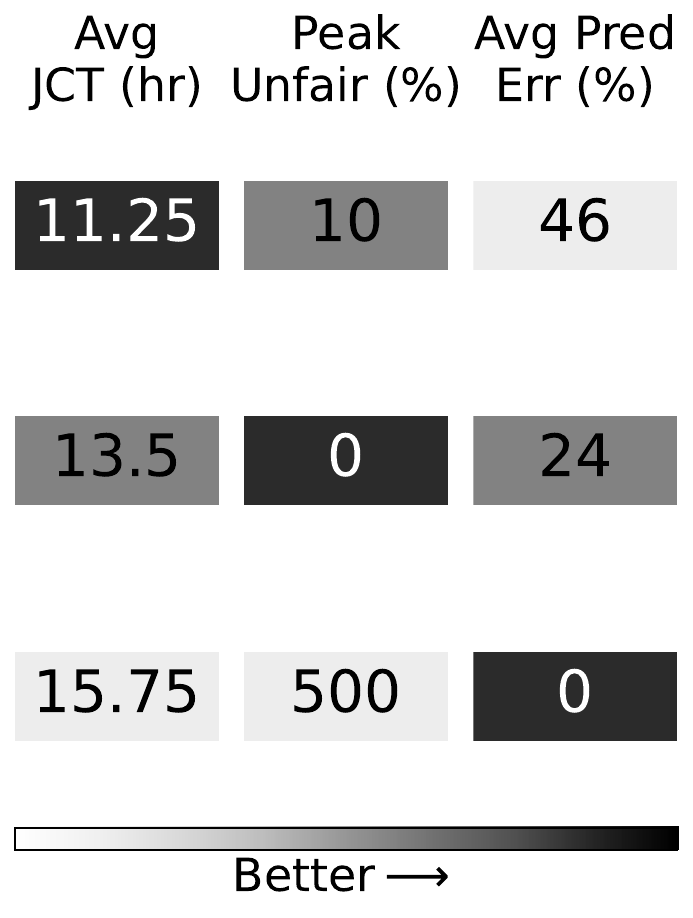}\label{subfig:toy-summary}}
    \caption{Toy example with 1 GPU, demonstrating the limitation of existing strategies. 
    (a) shows how the scheduling order changes as jobs arrive over time under the Tiresias~\cite{tiresias}, Themis~\cite{themis}, and FIFO~\cite{YARN} schedulers. Time moves from left to right with a new job arriving in each column. 
    The expected finish times for the current jobs are displayed above the current schedule. Jobs that are finished are grayed out.
    (b) summarizes the results for performance, fairness, and predictability for these  policies.}

    \label{fig:toy}
\end{figure*}

\subsection{Opportunities and Challenges}
Reliably predicting the completion time of a user's job requires multiple aspects of the cloud-ecosystem to be predictable.
For example, amongst other things it requires the knowledge of job sizes and the underlying scheduling system to be predictable~\cite{chronus}.
In this section, we highlight the opportunities that exist in making it feasible to predict job sizes for modern workloads and the challenge that arises when making reliable predictions for performance and fairness based schedulers used by the cloud today.

\paragraph{Opportunities: Exploiting workload characteristics.}
Predicting the completion time of a job requires knowing the job's size and its demand function (i.e., how its execution time will change based on the allocated resources).
Fortunately recent trends make this feasible.
\underline{(1) Intra-job predictability.}
A significant fraction of jobs that run in organizational clusters are DNN training and inference jobs~\cite{themis, philly, tiresias}.
These exhibit intra-job predictability; the time it takes to run an inference job~\cite{clockwork} or train a DNN for a specified number of epochs is fairly deterministic~\cite{themis}.
By profiling~\cite{AFS, pollux, gavel, resource-elastic-mlsys} or modelling~\cite{optimus, themis, shockwave,habitat, driple} the job's throughput and combining it with the provided job information (e.g., total number of epochs, convergence criteria) the size and the demand function can be estimated.
These and related techniques have been used beyond ML workloads as well; for requests in microservice deployments~\cite{grandslam}, compute jobs in data processing clusters~\cite{morpheus, cilantro} and I/O requests in storage clusters~\cite{mittos,linnos}.

\underline{(2) Recurring jobs.}
There is evidence that jobs repeat or exhibit similarity~\cite{corral, morpheus}.
For example, $60\%$ of training jobs exhibit DNN architecture similarity~\cite{modelkeeper}.
This can make history~\cite{3sigma} and sampling~\cite{slearn} based size estimation strategies highly effective.
Similar opportunities for determining the job size information exists in other contexts as well, such as network flow scheduling~\cite{flowsize-estimation}.
For the rest of the paper, we assume that job sizes can be determined using such techniques.

\paragraph{Challenge: Limitations of existing scheduling policies.} 
Despite knowing job size information, existing policies are limited in their ability to provide predictability -- the main reason is their use of preemption. 
Performance and fairness geared schedulers make heavy usage of \emph{unbounded preemption} by prioritizing and sharing jobs.
While preemption is central to achieving the desired objectives of these schemes, it can exacerbate unpredictability~\cite{morpheus}.
Under preemptive scheduling policies, future job arrivals can impact the completion times of existing jobs by preempting (expected) resources (e.g., GPUs) being used by them.
Preemption manifests in today's cloud systems in the following ways:

\noindent{}\underline{(1) Prioritization}. When a higher priority job arrives and needs to be scheduled, running jobs are paused or waiting jobs are pushed further back in the queue.
Several schedulers use prioritization to minimize JCTs and meet deadlines~\cite{tiresias, optimus, chronus, hypersched, pfabric, cheng2015resource,li2015dcloud}.

\noindent{}\underline{(2) Elastic Sharing}. Jobs may need to be multiplexed together as done by fairness and efficiency based schedulers~\cite{themis,gandiva,gandiva_fair,drf,AFS}.
A new job arrival can increase the multiplexing level and reduce the share of existing jobs, stretching their completion times~\cite{morpheus} or cause the scheduler to take away resources from existing less-efficient jobs and assign them to new jobs that can utilize the resource more efficiently~\cite{AFS, cilantro}.

On the other extreme are non-preemptive scheduling policies such as First-In-First-Out (FIFO) and reservation based schemes~\cite{philly} which are highly predictable as they guarantee resource allocation throughout the lifetime of a job --- future job arrivals do not impact current jobs in the system.
However, such schemes suffer from well known performance issues such as Head-Of-Line (HOL) blocking in the case of FIFO~\cite{2D, baraat, lps, AFS, tiresias} and poor utilization for reservation based schemes~\cite{antman, philly, PAI}.

\paragraph{Motivating example.}
We use a simple toy example (Fig.~\ref{subfig:toy-tiresias}) with four jobs (J1, J2, J3, and J4) to demonstrate these limitations.
We compare two popular GPU scheduling policies: Tiresias~\cite{tiresias} which prioritizes DNN training jobs with smaller remaining time and Themis~\cite{themis} which attempts to minimize peak unfairness with a FIFO scheduler~\cite{YARN}.
\footnote{We use a lease duration of 1 time unit for Themis and assume job size information is known for all schedulers}
Tiresias and Themis are representative of a large space of policies which either use size based scheduling or fair scheduling, respectively.
Figure~\ref{subfig:toy-summary} summarizes the results.
Unpredictability, is captured as: 
\begin{equation*}
    \text{\ecte{}} = \frac{JCT_{true} - JCT_{pred}}{JCT_{pred}}\%\\
\end{equation*}
while unfairness is captured as the additional time it takes for a job to complete compared to its fair finish-time~\cite{themis} (henceforth FFT) in percentage terms.

As new jobs arrive (moving left to right in Fig.~\ref{subfig:toy-tiresias}), both Tiresias and Themis result in a change in completion times of previous jobs.
For instance, in Tiresias (top row), when J2 and J4 arrive in the system (second and fourth column) there is an eight time unit increase in J1's predicted JCT each time.
While this approach achieves the minimum average JCTs, it results in the highest average prediction error --- 46\% \ecte{} in our example.
Similarly, in Themis (middle row), the scheduler's multiplexing of J1 and J2 causes J1's predicted completion time to increase by eight time units (second column).
While this ensures all jobs finish before their FFT (unfairness of 0\%) and also avoids HOL blocking, it has an avg \ecte{} of 24\%.
The FIFO scheduler (bottom row) achieves a prediction error of 0\% as it is non-preemptive.
However, it is the most unfair strategy (J3 has a JCT of 18) with the highest average JCTs because J2-J4 are smaller than J1 but have to wait behind.

We now discuss \sys{}, a generic resource scheduler that attempts to offer predictability while being flexible in balancing performance and fairness related objectives.
\section{Predictability Centric Scheduling (\sys{})}
\label{sec:mcs}

\textbf{Requirements.}
Our analysis in the previous section reveals that a scheduling policy with \emph{no preemption} (i.e., FIFO) results in maximum predictability.
However, this comes at a high cost in terms of performance (i.e., JCTs) and fairness, which makes it an \emph{impractical} option.
On the other extreme, there are scheduling policies that have \emph{unbounded preemption} (e.g., Fair-Share, Shortest Job First, etc.).
In these policies, an influx of future arrivals can arbitrarily stretch the completion time of an existing job, making them unsuitable for providing predictability.



This insight distills into the following two requirements that a scheduling policy must satisfy in order to provide predictability while being practical:

\begin{enumerate}
    \item[\circled{R1}] Predictability Requirement: a scheduling policy must have \emph{bounded preemption}.
    This is essential in order to provide reliable JCT predictions.

    \item[\circled{R2}] Flexibility Requirement: it should be able to approximate maximum predictability, optimal performance, and maximum fairness.
    Most importantly, it should be able to achieve Pareto-optimal trade-offs between these.
    This is essential for practicality.
    



    
\end{enumerate}


\begin{figure}[!t]
  \begin{center}
    \includegraphics[width=1\columnwidth]{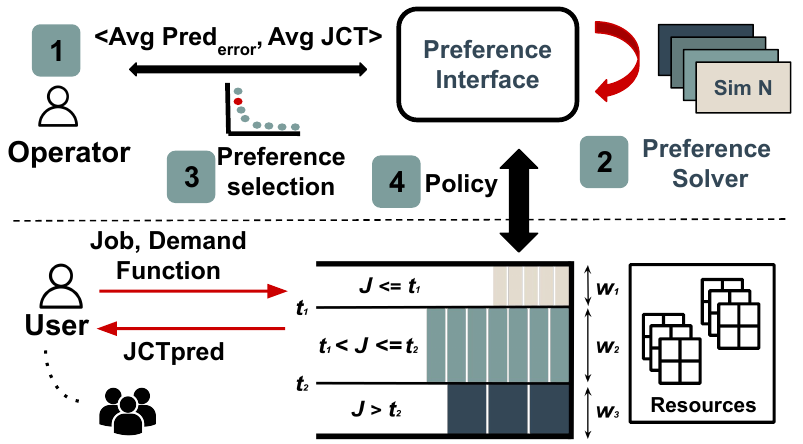}
  \end{center}
  \caption{Key components of \sys{}:
  The preference framework can be used by operators to specify high level objectives.
  The preference solver uses a simulation-based search strategy to find Pareto-optimal WFQ configurations that are then shared with the operator. 
  On the critical path, users submit their jobs along with the job's demand function and are given a \cte{}.}
  \label{fig:system}
\end{figure}

\paragraph{\sys{} Overview.}
Our solution to this end is \sys{}, a generic scheduling framework~(Fig.\ref{fig:system}), which captures these requirements using a high level preference interface~(\S\ref{subsec:preference_interface}), and meets them by using the well-known Weighted-Fair-Queuing (WFQ)~\cite{wfq-scott} policy in a novel way.
The inherent properties of WFQ, careful selection of various WFQ parameters (number of classes, weights, etc) along with how each job is mapped to a class and processed within it allow us to meet our objectives~(\S\ref{subsec:policy}).
Specifically, WFQ creates a fixed number of classes, assigns a guaranteed share of the resource capacity (class weights) to each class and schedules jobs within a class in FIFO order -- this allows WFQ to satisfy our predictability requirement (R1).
Similarly, the number of classes, class weights, and how jobs are mapped to each class are \emph{tunable} decisions, allowing it to offer the desired flexibility (R2).

A key component of \sys{} that enables the above functionality is the \emph{preference solver} (\S\ref{subsec:preference_solver}), which translates the specified high level objectives into Pareto-optimal WFQ configurations using a simulation-based search strategy.
It does this as coarser timescales to capture changes in cluster deployment and workload changes.
While the configuration space is large, we use an intelligent parameterization of WFQ (e.g., limiting coefficient of variation among job sizes within a class) making this process feasible.

Finally, an important benefit of \sys{} is that it is a generic scheduling policy -- it operates on the notion of a \emph{job} which could be a network flow or a DNN training job etc.
To deal with the varying needs of these different scenarios, in \sys{}, a job is defined using a demand function.
The demand function is a mapping between the job's execution time and the resources allocated to it i.e., $demand(n) \mapsto{} T_{exec}$ and has a minimum ($demand_{min}$) and maximum ($demand_{max}$) resource allocation bound, denoting the execution time under the lowest and highest possible allocation.
For scenarios like network (co)flow scheduling~\cite{baraat, 2D, aalo}, the demand function is simple and could be specified as the (co)flow size/allocated bandwidth.
For the DNN scheduling scenario which could have GPUs with different speedups and affinity, the demand functions could be more sophisticated (e.g., because of sublinear scaling), as discussed in \S\ref{sec:system}.


We now explain in detail, our choice of using WFQ as a building block~(\S\ref{subsec:policy}), followed by preference solver and interface.

\subsection{WFQ under \sys{}}\label{subsec:policy}
We begin by motivating why WFQ is a useful starting point and then share \sys{}'s careful usage of WFQ in meeting our objectives.
Our observation is that a lack of preemption as in FIFO and a non-zero guaranteed resource share for jobs is crucial for predictability.
WFQ uses FIFO scheduling within each class and across classes the resources are shared according to, strictly positive, class weights, helping us satisfy the predictability requirement.
To highlight the flexibility of WFQ, we show how it can be configured to optimize for extreme points in the trade-off space of maximum predictability, performance and fairness.

\begin{itemize}[leftmargin=*]
    \item \textbf{Maximum predictability:}
    WFQ with a single class is exactly FIFO scheduling which achieves a prediction error of 0\%
    
    \item \textbf{Near-optimal performance:}
    Shortest Job First (SJF) is near-optimal in minimizing avg JCTs for a single bottleneck~\cite{srsf-optimal}.
    WFQ can map each job to its own class and give a higher weight to classes with smaller jobs, approximating SJF as shown by prior work~\cite{PWFQ, aalo}.

    \item \textbf{Max-Min Fairness:}
    If each job is mapped to its own class and each class gets an equal weight, WFQ can emulate Max-Min fair allocation which minimizes unfairness for a single bottleneck~\cite{choosy}.
\end{itemize}
As our analysis in~\S\ref{sec:case} reveals, a combination of these objectives is more practical.
WFQ offers the necessary baseline flexibility in the class creation, job mapping and weight assignment strategy.
This motivates that we can achieve a combination of these objectives as well, which leads to \sys{}'s preference interface~\S\ref{subsec:preference_interface}.

\paragraph{Beyond vanilla WFQ.}
Our core idea is the novel use of WFQ to meet our objectives.
First, \sys{} intelligently chooses the number of classes, class weights and the job-to-class mapping strategy to find various Pareto-optimal configurations, including extreme points, such as FIFO, SJF and Max-Min Fair Share.
In \sys{} jobs are mapped to different classes based on their size and a set of thresholds ($t_i's$), while strictly positive class weights ($w_i$'s) dictate the guaranteed resource share for each class.
For example, jobs with size $>t_k$ and $\leq t_{k+1}$ will be mapped to the $k^{th}$ class.

Second, within a class, \sys{} deviates slightly from a strict FIFO schedule in favor of improving performance and fairness.
In \sys{} a job's demand function is used to cap the resources allocated to it.
For example, a job in front of its class may not be assigned all of the guaranteed resource share of its class (as in strict FIFO), and instead some of the resources may be allocated to the jobs behind it.
Doing this becomes important when catering to sophisticated job demand functions (e.g., if a job's speedup exhibits diminishing returns w.r.t increase in allocated resources) such as those in the context of DNN jobs~(\S\ref{sec:system}).
Finally, to ensure work-conservation, any residual allocation is then redistributed first within a class in FIFO order by incrementally relaxing the cap on each job's demand function and then across classes proportional to their weights.
We expose the class weights, thresholds and the demand capping criteria to the preference solver which searches over the space of possible choices of these parameters in order to discover Pareto-optimal configurations~(\S\ref{subsec:preference_solver}).

\paragraph{A note on prediction error.}
Since \sys{} is work-conserving, a job may get a higher resource share compared to its guaranteed share.
For example, if a job arrives when no other jobs are present, it will be allocated all the available resources.
This may lead to prediction errors if jobs of other classes arrive in the future.
To mitigate sudden and drastic changes in class occupancy, we exploit the fact that cloud systems are typically highly loaded~\cite{tiresias}.
If we avoid creating too many classes, we can increase the likelihood of all classes being occupied by some jobs.
Furthermore, the exact load of a class can be controlled by the class thresholds and weight assignment strategy.
These observations guide us in discovering Pareto-optimal WFQ configurations.

\begin{figure}[!t]
    \centering
    \subfloat[width=0.5\columnwidth][Trace \#6214e9]{\includegraphics[width=0.495\columnwidth]{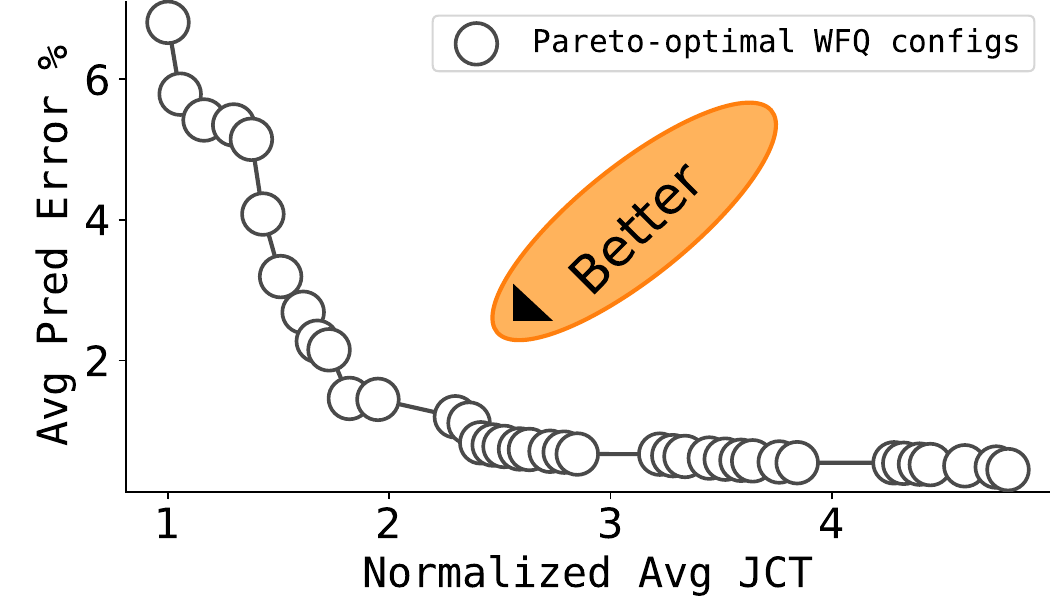}\label{fig:pareto_avg_6214e9}}
    \subfloat[width=0.5\columnwidth][Trace \#ee9e8c]{\includegraphics[width=0.495\columnwidth]{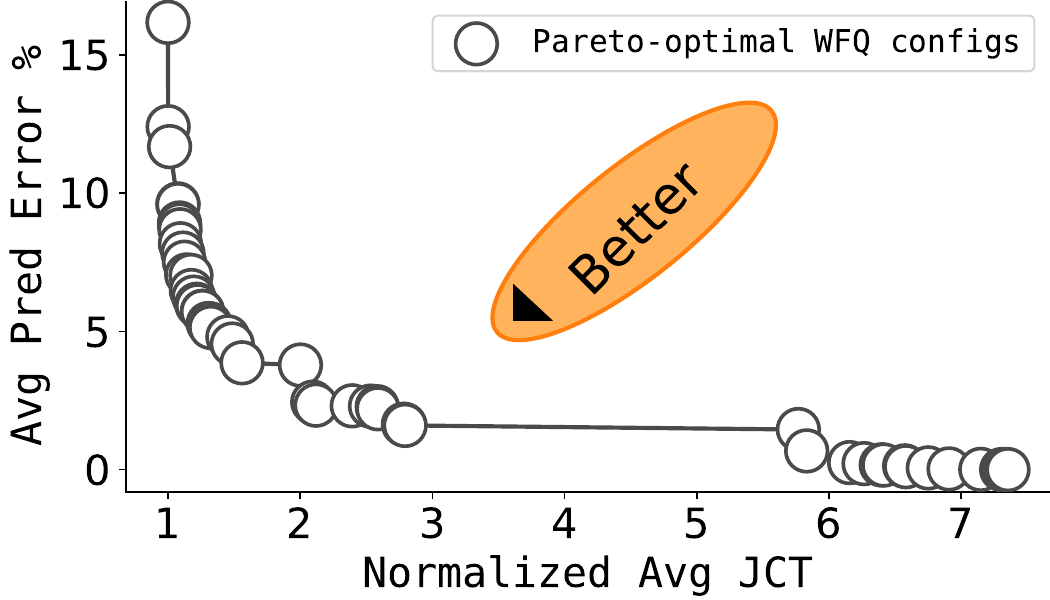}\label{fig:pareto_avg_ee9e8c}}
    \caption{ 
    Pareto front of the trade-off between \ecte{} and normalized average JCT for workload-2~(\S\ref{sec:eval}).
    \emph{Better} indicates WFQ configurations that achieve a tight bound on average/tail \ecte{} while incurring the smallest possible increase in average JCT.
    }
        
    
    \label{fig:pareto}
\end{figure}

\subsection{Preference Interface}
\label{subsec:preference_interface}
\sys{} exposes a simple yet expressive bi-directional interface that allows operators to specify high level objectives and present Pareto-optimal trade-offs (WFQ configurations) to choose from.
This is unlike other tunable systems~\cite{gavel, decima, selftune} which assume operators are aware of the trade-offs involved --- i.e., \sys{} \emph{actively} tries to help the operator in making an informed choice.
Our decision to use Pareto-optimal choices as a way to support informed decision making is grounded in fundamental literature on multi-objective decision making~\cite{pareto-front-dm, pareto_multi_objective}, which maps neatly to the problem \sys{} is trying to address: predictability while being practical.

The preference interface itself, is general enough to be used in scenarios beyond predictability as well.
For example it can be used to strike a balance between fairness and performance (e.g., Carbyne~\cite{carbyne}) and between minimizing average and tail JCTs~\cite{carbyne, 2D, baraat, lps}.
The preference interface exposes the following API:


\noindent\makebox[\columnwidth]{\rule{\columnwidth}{0.2pt}}
\begin{lstlisting}[language=C, 
basicstyle=\ttfamily, keywordstyle=\color{blue}]
void SetPreference(
    Obj1 <Metric, Measure>,
    ...,
    ObjN <Metric, Measure>
);
List<WFQConfig> GetParetoFront();
void SetWFQConfig(WFQConfig config);
\end{lstlisting}
\noindent\makebox[\linewidth]{\rule{\columnwidth}{0.2pt}}

\noindent{}The current \sys{} API supports three \texttt{Metrics}: Performance (\textit{JCT}), Fairness (\textit{unfairness}), and Predictability (\ecte{}).
The \texttt{SetPreference()} method is used to specify the list of objectives; repeated entries are allowed to support exploring trade-offs across \emph{different} measures of the \emph{same} metric.
For each objective, \texttt{avg(.)} or a particular \texttt{percentile(p)} needs to be specified as a \texttt{Measure}.
The \texttt{GetParetoFront()} can be invoked to get the set of Pareto-optimal WFQ configurations.
Finally, the \texttt{SetWFQConfig()} method can be used to select a specific \texttt{WFQConfig}.

We now show how the API was used to target scenarios covered in our evaluation.

\paragraph{Average JCTs \emph{vs} Average \ecte{}:}
Minimizing average JCTs is a popular performance objective and has been a focus of several scheduling policies~\cite{tiresias, optimus}.
To explore the trade-off between performance and predictability, one can specify it as \texttt{SetPreference(<JCT, avg>, <\ecte{}, avg>)}.
We use this for evaluating \sys{} for workload-1 and workload-2 in~\S\ref{sec:eval}.

\paragraph{Average JCTs \emph{vs} Tail \ecte{}:}
Prediction error can be tightly bound by specifying the tail \ecte{} (e.g., p99) as a measure of predictability.
In such a case, the objectives would stay the same as in the above example, however, the measure for \ecte{} would change from \texttt{avg(.)} to \texttt{percentile(99)}.
\sys{} uses this specification for workload-3 where low p99 \ecte{} is challenging to achieve with other policies.
        

\paragraph{Pareto fronts.}
Figure~\ref{fig:pareto} shows the Pareto-optimal space shared by \sys{} with the operators for realistic DNN training workloads. 
Each point is a WFQ configuration that makes a certain trade-off.

\subsection{Preference Solver}
\label{subsec:preference_solver}
The preference solver is responsible for finding Pareto-optimal WFQ configurations for the objectives specified.
It uses a (plug-and-play) multi-objective search algorithm to navigate the space of possible WFQ configurations.
The optimization parameters consist of the number of classes, class weights, class thresholds, and the resource allocation cap.
Finding Pareto-optimal configurations in this setting is challenging due to the combinatorial nature of the configuration space.
The solver intelligently parameterizes each configuration to make the search process feasible.
The solver uses a simulation-based approach to evaluate the performance, predictability and fairness of a particular WFQ configuration.
These are fed to the search algorithm which decides the configurations to keep, try out next and discard.

\paragraph{Intelligent parameterization.}
To reduce the number of optimization parameters we use the following heuristics:
\begin{itemize}[leftmargin=*]
\item \textbf{Creating classes and thresholds:}
Large variation in job-sizes within a class could lead to HOL blocking but creating too many classes increases preemption events and deteriorates predictability.
Instead of an arbitrary number of classes, we create them based on the squared coefficient of variation ($C^2$) in the job-sizes within a class~\cite{2D}.
We use a tunable parameter $\mathcal{T} > 0$ to ensure that classes are created such that $C^2$ of job-sizes within each class is $\leq \mathcal{T}$.

\item \textbf{Systematic class weight selection:}
Higher weights given to classes with smaller jobs improves performance for most workloads.
On the other hand, a balanced weight assignment strategy may improve fairness instead.
Based on this, we constrain class weight for the $i^{th}$ class such that $w_i = e^{-i\times{}\mathcal{W}}$.
$\mathcal{W}$ is a tunable parameter which controls the relative weights for each class.
For heterogeneous deployments, containing several resource types (e.g., $k$ different GPU types) we use $\mathcal{W}_{1} \hdots{} \mathcal{W}_k$.

\item \textbf{Finding demand caps:}
The resource efficiency of a job is used to decide its allocation cap and it is computed as $\zeta{}(n) = \frac{demand_{min}}{n\times{}demand(n)}$, where $demand_{min}$ is a job's execution time under its minimum possible allocation (e.g., 1GPU) and it is a non-increasing function of the allocated resources.
For linear scaling jobs, $\zeta{} = 1$, while for jobs that scale sublinearly, $0<\zeta{}\leq{}1$.
Instead of finding an optimal allocation cap for each job, we introduce a global tunable threshold $\zeta{}_{min}$.
Using this, a job's resource allocation is capped at $k$ such that $\zeta{}(k) \geq{} \zeta_{min}$.
Intuitively, a low (high) $\zeta{}_{min}$ means the scheduler is more (less) tolerable to inefficient jobs.
\end{itemize}
Using these heuristics, WFQ$(\mathcal{T}, \mathcal{W}, \zeta{}_{min})$ becomes the succinct parameterization of each configuration.
Different values for these parameters results in different trade-offs between the objectives specified by the operator.
Next we discuss our simulation based search strategy to discover Pareto-optimal choices for these parameters.


\paragraph{Simulation-based search.}
We use a simulation based approach to discover Pareto-optimal WFQ configurations.
We use a high fidelity event-based simulator which takes as input a WFQ configuration i.e., a $(\mathcal{T}, \mathcal{W},\zeta{}_{min})$, evaluates it for a statistically significant number of job arrivals ($\approx 1000$) and outputs the resulting JCT, FFT and \cte{} for each job.
The results are then fed to the search algorithm.


The search algorithm samples the search space of possible WFQ configurations and interacts with the simulator to converge to Pareto-optimal solutions.
As such, this can be any off-the-shelf multi-objective optimization algorithm.
We use SPEA2, which is a popular multi-objective optimizer based on evolutionary search~\cite{spea2}.

While we don't have any theoretical basis for the convergence properties of our approach, it works well in practice and can timely ($\approx 1$hr) discover the Pareto front for a reasonably sized GPU cluster.
Our evaluation confirms that Pareto-optimal configurations found using simulations follow the same trade-offs on the testbed experiment~(\S\ref{subsec:eval-testbed}).
We micro-benchmark the feasibility of the simulation-based search strategy in~\S\ref{subsec:micro}.

\section{\sys{} for GPU scheduling}\label{sec:system}
We now describe the realization of \sys{} for DNN scheduling on GPU clusters, 
highlighting important differences and how our abstraction of a job and demand function handles these differences.

\paragraph{DNN jobs.}
The notion of a job is either a single DNN training job or a collection of DNN trials being run as part of a hyper-parameter tuning task (i.e., AutoML).
The demand function for such workloads can be complicated.
Modern DNNs require distributed training (e.g., data parallelism) on multiple GPUs.
They are known to have sublinear speedup w.r.t to the (1) number, (2) type and (3) locality of GPUs allocated to them~\cite{AFS, pollux, themis, gavel}.
\sys{} relies on existing techniques, such as throughput modelling and profiling, to estimate a job's demand function.
As described in \S\ref{sec:mcs}, the demand function describes how the job's execution time changes with different resource allocations.
Since allocations have three dimensions: locality, GPU type and number of allocated GPUs, the demand function takes as input different combinations and returns the corresponding execution time.
This is akin to the notion of bids in Themis~\cite{themis} and throughput in Gavel~\cite{gavel}.

\paragraph{Role of demand functions.}
\sys{} uses a job's demand function to map jobs to their respective class and to determine the resource allocation cap of jobs \emph{within} a class.
We use $demand_{min}$ as the definition of a job's size and use it to map jobs to classes based on each job's $demand_{min}$ and the class thresholds ($t_i's$).
Capping resource allocation is based on the insight that for several DNN models, allocating GPUs up to the job's maximum demand ($demand_{max}$) may be sub-optimal.
For such jobs, max demand is capped to improve performance and fairness while controlling predictability --- in a conservative way compared to other schemes such as AFS~\cite{AFS}.
We evaluate this approach and show that it works for DNN workloads consisting of jobs that scale sub-linearly~(\S\ref{subsec:eval-sim}).
As described in~\S\ref{subsec:preference_solver}, the allocation cap is a tunable parameter for the preference solver and can be adjusted for different trade-offs and workloads.

\paragraph{Resource heterogeneity.}
To handle resource heterogeneity we reuse an existing solution: Gavel~\cite{gavel}.
Gavel makes a scheduling policy heterogeneity-aware.
It supports hierarchical policies with weighted-fairness across entities and FIFO scheduling within an entity.
Our observation is that \sys{} is ``Gavel-friendly''.
The different parameters of WFQ (e.g., number of classes, weights etc.) map elegantly to the primitives provided by Gavel.
Once an operator chooses the Pareto-optimal configuration, \sys{} can convert it into an optimization problem that Gavel can solve for.

\paragraph{Implementation.}
We implement \sys{} as a central coordinator in Python and use the Ray cluster manager~\cite{ray} for GPU allocation enforcement as well as for general cluster management tasks such as fault tolerance.
Each job is either a single trial or consists of multiple trials as part of a hyperparameter tuning algorithm provided by RayTune.
We use a custom \texttt{ray\_trial\_executor} to control starting, stopping and preempting individual trials based on the allocations computed by \sys{}.
To determine the remaining service requirements of running jobs, we use various callbacks (e.g., \texttt{on\_step\_start}) exposed by Tune to get the exact number of iterations completed by each job and multiply it with the profiled time per iteration.

In addition to the central coordinator, \sys{} consists of an agent, which uses information about running jobs to provide a prediction interface.
This interface returns a \cte{} in real time to the user whenever they submit their jobs.
\sys{} agent computes this \cte{} by ``virtually'' playing out (i.e., in a simulated setting) the current snapshot of the cluster state (e.g., running jobs, available GPUs etc.) to determine the time at which the job will end.
It accounts for preemption, restart overheads and GPU allocation to training throughput relationships of all current jobs based on measurements collected during the initial profiling phase.
This approach is inspired by prior work~\cite{jockey,predict-mapreduce}, which use a simulator to compute a job's duration under different resource allocation strategies.

\section{Evaluation}\label{sec:eval}
We evaluate \sys{} on a 16 GPU cluster with a realistic AutoML style workload to validate our observations.
We also cover additional workloads at a larger scale using an event-based simulator.
Our evaluation covers different application workloads (e.g., heavy-tailed vs. light-tailed, AutoML apps vs single DNNs), different scheduling schemes (e.g., Tiresias~\cite{tiresias}, Themis~\cite{themis}) and different metrics (e.g, Avg, p99).

Our evaluation attempts to answer the following key questions:

\begin{itemize}[leftmargin=*]
    \item \textbf{How does \sys{} perform in terms of \ecte{} compared to other schemes?}
    Our testbed results reveal that \sys{} configurations achieve significantly lower \ecte{} ($20\%$) while being within 10\% of high performing schemes on the performance side.
    
    \item \textbf{Does \sys{} work well across different workload types?}
    The flexibility and predictability provided by \sys{} holds across different workloads and across preference specifications.
    \sys{} can discover configurations can bound the tail \ecte{} to be within $100\%$ compared to AFS~\cite{AFS} and Tiresias~\cite{tiresias} which suffer from $\geq 300\%$ error at the tail.
    
    \item \textbf{Are \sys{} configurations fair?}
    \sys{} configurations that are optimized for the performance vs \ecte{} trade-off do not necessarily suffer from unfairness because each class is guaranteed a GPU share which helps in avoiding starvation.
        
    \item \textbf{Is the search process feasible?}
    Our microbenchmark reveals that the search process can complete within O(hr), making it practical to use and \sys{} configurations discovered using the simulation based search-strategy observe the same trade-off \emph{trends} on the actual testbed.
    
\end{itemize}

\subsection{Experimental Setup} \label{subsec:experimental-setup}

\paragraph{Testbed.}
Our testbed cluster consists of 16 1-GPU c240g5 machines in the Wisconsin Cloudlab cluster~\cite{cloudlab}.
Each machine has one NVIDIA P100 GPU with 12GB GPU memory.

\paragraph{Simulation.}
We use an event-based simulator to cover workloads that contain jobs requiring O(100) GPUs on a homogeneous 64 GPU cluster.
We have verified the fidelity of our simulator with trace results from Microsoft~\cite{philly} and our testbed results with the difference being within 5\%.


\begin{table}[]
\resizebox{\columnwidth}{!}{%
\begin{tabular}{@{}cccc@{}}
\toprule
         & Testbed (16 GPUs) & \multicolumn{2}{c}{Simulations (64 GPUs)} \\ \midrule
Workload & Workload-1        & Workload-2          & Workload-3          \\
Job Type & AutoML            & DNN                 & DNN                 \\
DNN/job  & 1-20              & 1                   & 1                   \\
GPUs/DNN & 1                 & 1-52                 & 1-8                \\ \bottomrule
\end{tabular}%
}
\caption{Summary of the settings used to evaluate \sys{}}
\label{table:eval-table}
\end{table}

\paragraph{Pareto search.}
The Pareto-optimal configurations for our workloads are discovered by the preference solver \S\ref{subsec:preference_solver} running on a cluster of 10 c220g5 machines in the Wisconsin Cloudlab cluster~\cite{cloudlab}.
It is important to note that these configurations are discovered and evaluated on different sampled subsets of the workload i.e., there exists a notion of training set vs testing set.

\paragraph{Workloads.}
Table~\ref{table:eval-table} summarizes the characteristics of our candidate workloads. We now discuss these workloads in detail.
\begin{itemize}[leftmargin=*]
    \item \textbf{Workload-1:} We borrow this workload from Themis~\cite{themis} (referred in their work as \emph{Workload-1}).
    For our testbed evaluation we use scale down the maximum number of trials per app to 20 and the maximum service time to 2 GPU-hours,
    The maximum number of GPUs per trial is set to 1.
    Each trial tunes a different hyper-parameter (learning rate and momentum) of popular vision models from the VGG family~\cite{vgg}.

    \item \textbf{Workload-2:} We use 5 traces (\textsf{0e4a51}, \textsf{b436b2}, \textsf{6214e9}, \textsf{6c71a0}, \textsf{ee9e8c}) from the Philly cluster~\cite{msr-fiddle} containing the largest number of jobs.
    In contrast to other workloads, jobs in these traces exhibit sub-linear scaling.
    We use the scaling data shared by Hwang et al. on Github~\cite{sched-sim}.

    \item \textbf{Workload-3:} This is borrowed from Gavel~\cite{gavel} (referred in their work as \emph{continuous-multiple}).
    
    
\end{itemize}
A common characteristic of these workloads is that the minimum requirement of any job is 1 GPU i.e., as long as there is at-least one GPU available, a job can start.
This also holds true for RayTune apps which we use in our testbed evaluation.

\paragraph{Scheduling policies.} We compare \sys{} against FIFO and recently proposed GPU scheduling systems (Themis, Tiresias, AFS) for different criteria.
All scheduling policies considered in our evaluation are ``work-conserving'' and elastic i.e., they redistribute unused GPUs amongst other jobs according to the policy.
For example for FIFO if a job only needs $k$ GPUs and $n$ are available, $n-k$ are attempted to be allocated to the next-in-line jobs.

We now describe our implementation of Themis, AFS, and Tiresias that we use in our evaluation. 
\begin{itemize}[leftmargin=*]
    \item \textbf{Themis}~\cite{themis}.
    On every resource change event and lease duration expiry, in-progress jobs report their fair-finish-time and we allocate GPUs to jobs in descending order of the reported number.
    We do not consider the scenario where jobs could lie and thus do not require the partial allocation mechanism.
    The lease duration is set to 10 minutes as per the author instructions.
    
    \item \textbf{Tiresias}~\cite{tiresias}.
    Since we assume complete knowledge about job sizes, here Tiresias emulates the Shortest-Remaining-Service-First (SRSF) policy.
    \footnote{Referred to as Tiresias-G in their paper~\cite{tiresias}}
    As such, GPUs are first allocated to jobs with the lowest remaining service times on every resource change event.
    
    \item \textbf{AFS}~\cite{AFS}. This scheduler tries to minimize avg and tail JCTs while maximizing resource efficiency.
    On every resource change event we compute each job's allocation using the \textsc{AFS-L} algorithm shared by the authors.
\end{itemize}

\paragraph{\sys{} configurations.}
We use three configurations for \sys{}: (1) \sys{}-pred, (2) \sys{}-JCT, and (3) \sys{}-balanced.
Each configuration is making a different trade-off compared to each other.
For example, \sys{}-pred has the highest JCT but the lowest \ecte{} amongst the three.
For each workload and objective the exact WFQ configuration is different.

\paragraph{Comparison criteria.}
We evaluate the merit of \sys{} on three fronts:
\begin{enumerate}[leftmargin=*]
    \item Job Completion Times (JCTs): A commonly used metric to evaluate the performance of scheduling policies.
    \item Unpredictability (\ecte{}): A proxy to capture the error in \cte{}.
    \item Unfairness: It captures the extra time taken by a job to complete, compared to its fair-finish-time (FFT) and is 0\% for jobs that complete before their FFT.
\end{enumerate}
We consider all important statistics such as the average and tail (e.g., p99 \ecte{}, avg JCTs) for all objectives.
For each objective, a lower value is better.
\footnote{The testbed result is an average across 3 seeds while simulation results are an average across 5 seeds}



\begin{figure}[!t]
    \centering
    \includegraphics[width=1\columnwidth]{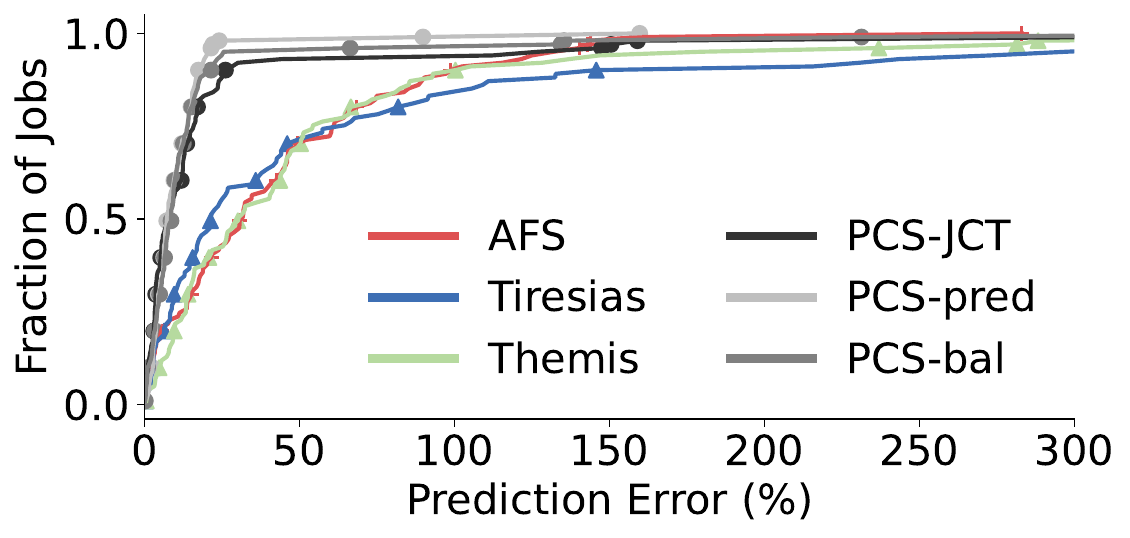}
    \caption{[\textsc{Testbed}] Distribution of \ecte{} showcasing three configurations of \sys{} discovered by \sys{} --- performance oriented, predictability oriented and balanced compared to other schemes.}
    \label{fig:testbed-pred}
\end{figure}

\begin{figure}[!t]
    \centering
    \subfloat[][JCTs]{\includegraphics[width=0.5\columnwidth]{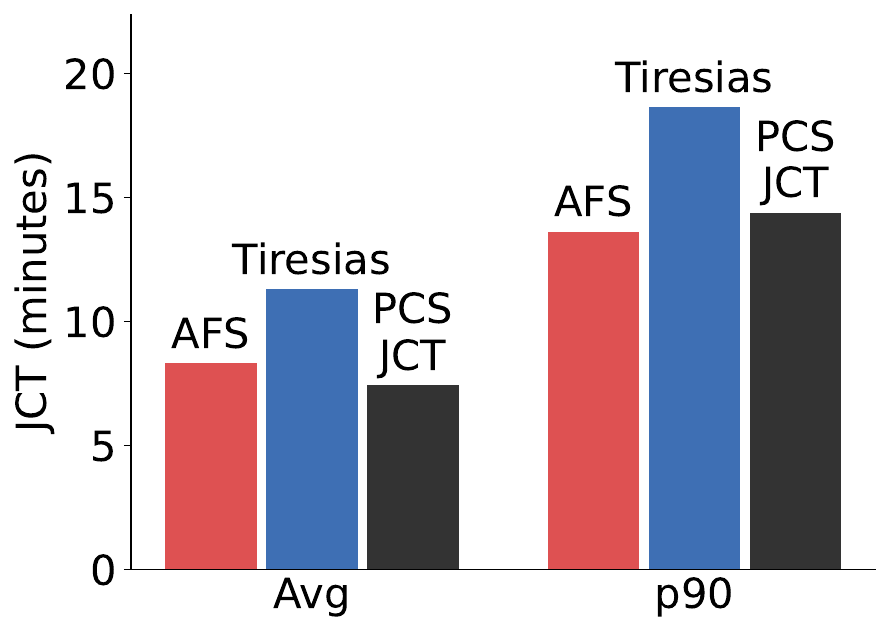}\label{subfig:testbed-act}}
    \subfloat[][\ecte{}]{\includegraphics[width=0.5\columnwidth]{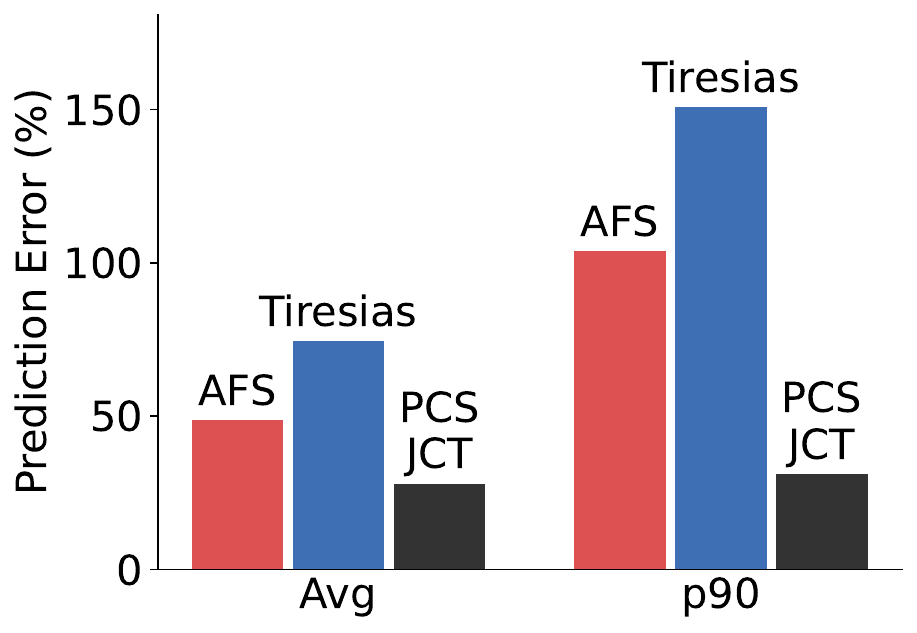}\label{subfig:testbed-act-error}}
    \caption{[\textsc{Testbed}] Zooming into the trade-off between performance and predictability. \sys{} is within 1.1$\times{}$ AFS at p90 JCT, with significant improvement to predictability.}
    \label{fig:testbed-perf-vs-pred}
\end{figure}

\begin{figure}[!t]
    \centering
    \subfloat[][Distribution of unfairness]{\includegraphics[width=0.5\columnwidth]{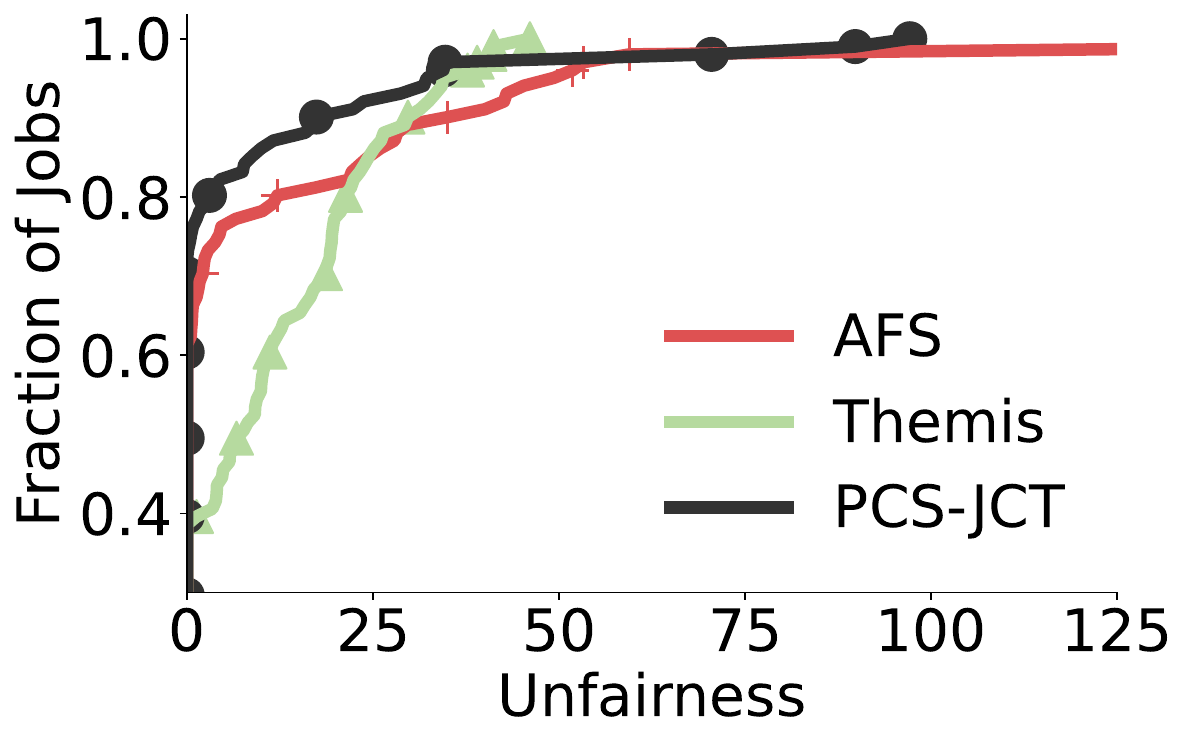}\label{fig:testbed-unfairness}}
    \subfloat[][Pareto font]{\includegraphics[width=0.5\columnwidth]{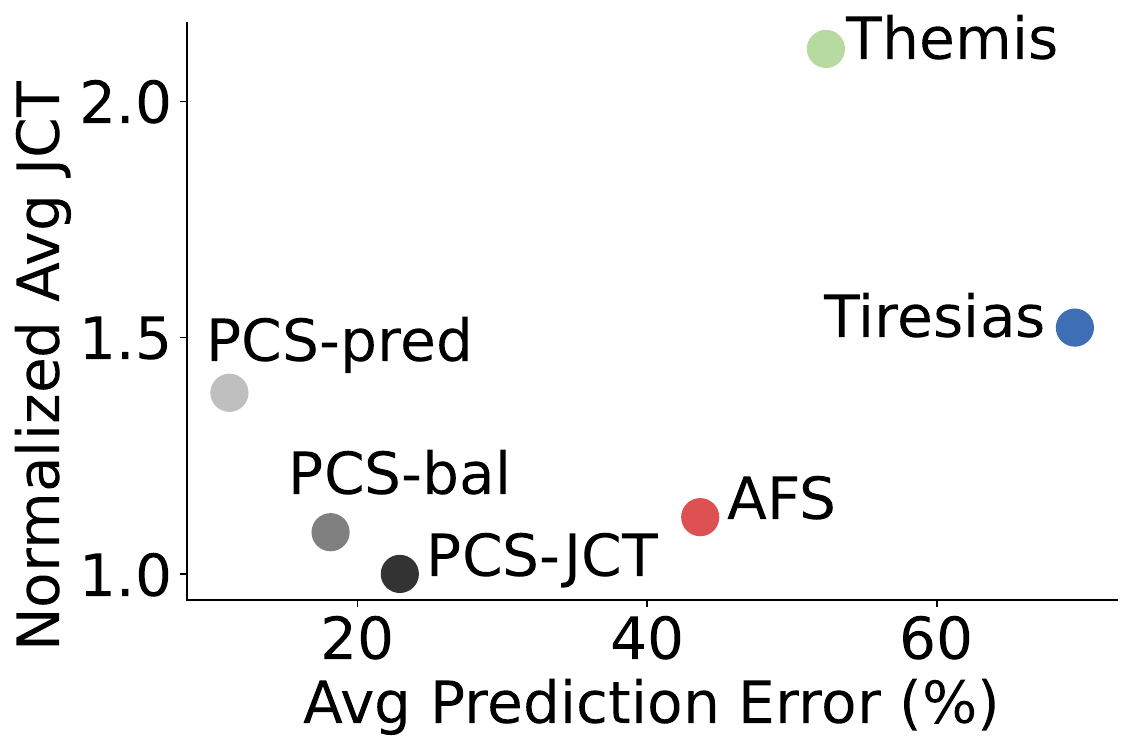}\label{fig:testbed-pareto-avg-avg}}
    \caption{[\textsc{Testbed}] (a) shows the CDF of unfairness showcasing that \sys{} does not significantly compromise on fairness compared to a policy that optimizes for it.
    (b) highlights the Pareto-optimal configurations discovered in a simulated environment observe the same trend on the testbed evaluation.}
\end{figure}

\subsection{Testbed Experiment}
\label{subsec:eval-testbed}
For our main experiment we compare three \sys{} configurations, discovered by the preference solver workload-1 against other schemes.

\paragraph{A tight bound on \ecte{}.}
Figure~\ref{fig:testbed-pred} shows the CDFs of \ecte{} achieved by different scheduling schemes and the three \sys{} configurations.
We observe that all \sys{} configurations are able to achieve significantly lower \ecte{}.
At p90, the difference is an $80\%$ lower error achieved by all configurations compared to other schemes.
At higher percentiles, \sys{}-pred provides the lowest worst-case \ecte{} of $150\%$ while other schemes have a long tail.
\sys{}-JCT still has a lower \ecte{} up until p95.

\paragraph{Negligible performance sacrifice for high predictability.}
Figure~\ref{fig:testbed-perf-vs-pred} zooms into the performance versus predictability trade-off achieved by \sys{}-JCT compared to AFS and Tiresias which aim to minimize JCTs.
We see that \sys{}-JCT achieves equivalent performance to AFS and Tiresias for the average JCTs.
It is within 1.1$\times{}$ of AFS at p90, however this trade-off results in significant improvement on the predictability front, where Tiresias and AFS suffer.
\ecte{} under \sys{}-JCT is within $20\%$ for average and p90 \ecte{} while AFS and Tiresias have $\geq 40\% (\geq 100\%)$ prediction error at the average (p90).
This signifies that \sys{}-JCT trades off negligible performance to significantly improve predictability.
Another source of improvement we observe is that since \sys{} makes limited use of preemption, overheads associated with preempting running jobs are reduced compared to other schedulers.

\paragraph{Unfairness.}
Figure~\ref{fig:testbed-unfairness} compares the unfairness for \sys{}-JCT compared to AFS which optimizes for average JCT and Themis which minimizes unfairness.
\sys{} achieves lowest unfairness till p95 and has the worst-case unfairness $\leq{}100\%$ compared to AFS which has a worst-case unfairness $>200\%$.
Not surprisingly, Themis offers the tightest bound on the worst-case unfairness of less than $50\%$. 

\paragraph{Pareto-optimality.}
Finally, figure~\ref{fig:testbed-pareto-avg-avg} shows different \sys{} configurations that achieve different trade-off points in the space of avg JCT vs avg \ecte{}.
As expected, \sys{}-JCT has the lowest avg JCT, while \sys{}-pred achieves the lowest average \ecte{}.

\subsection{Simulation Experiments}
\label{subsec:eval-sim}

We now consider different workloads at a larger scale in simulations and show the trade-offs achieved by suitable \sys{} configurations compared to performance and fairness optimal schedulers.


\begin{figure}[!t]
    \centering
    \subfloat[][Normalized avg JCT]{\includegraphics[width=\columnwidth]{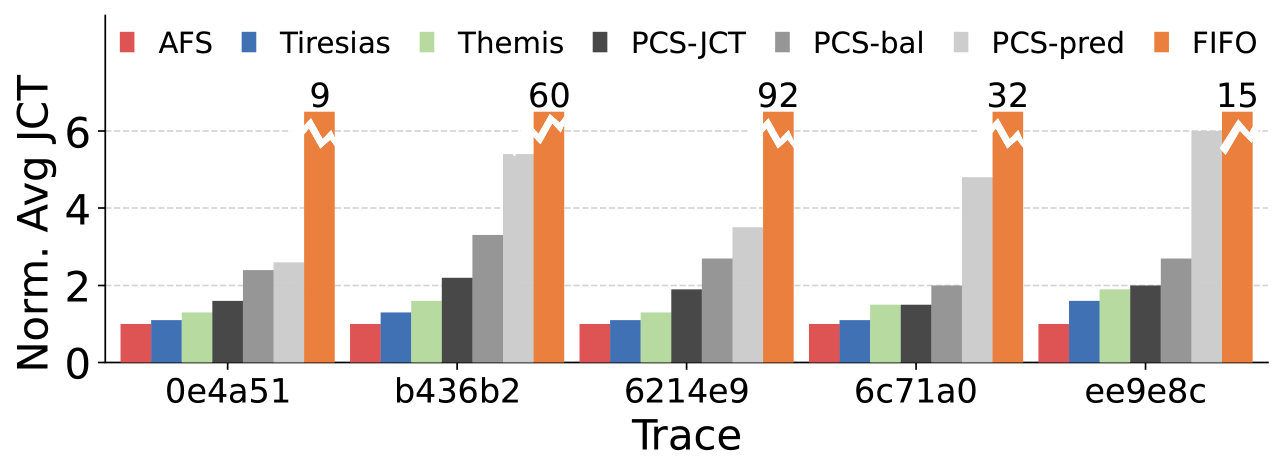}\label{subfig:traces-jct}}
    \vspace{5pt}
    \subfloat[][Predictability]{\includegraphics[width=\columnwidth]{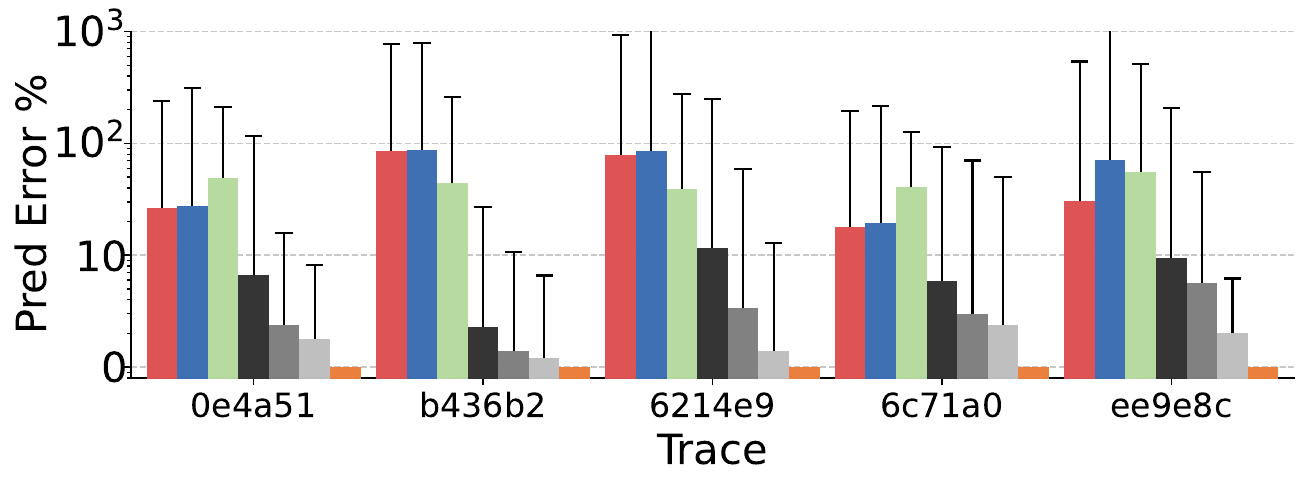}\label{subfig:traces-error}}
    \caption{[\textsc{Sim}]: \sys{} for workload-2. a) Most \sys{} configurations are within 1.5-4$\ttimes{}$ of the performance optimal policies while b) shows that they drastically reduce the average and tail \ecte{}. In b), the bar height (line) represents average (p99) \ecte{} and the y-axis follows a logscale.}
    \label{fig:traces}
\end{figure}

\begin{figure}[!t]
    \centering
    \subfloat[][Prediction Error]{\includegraphics[width=0.5\columnwidth]{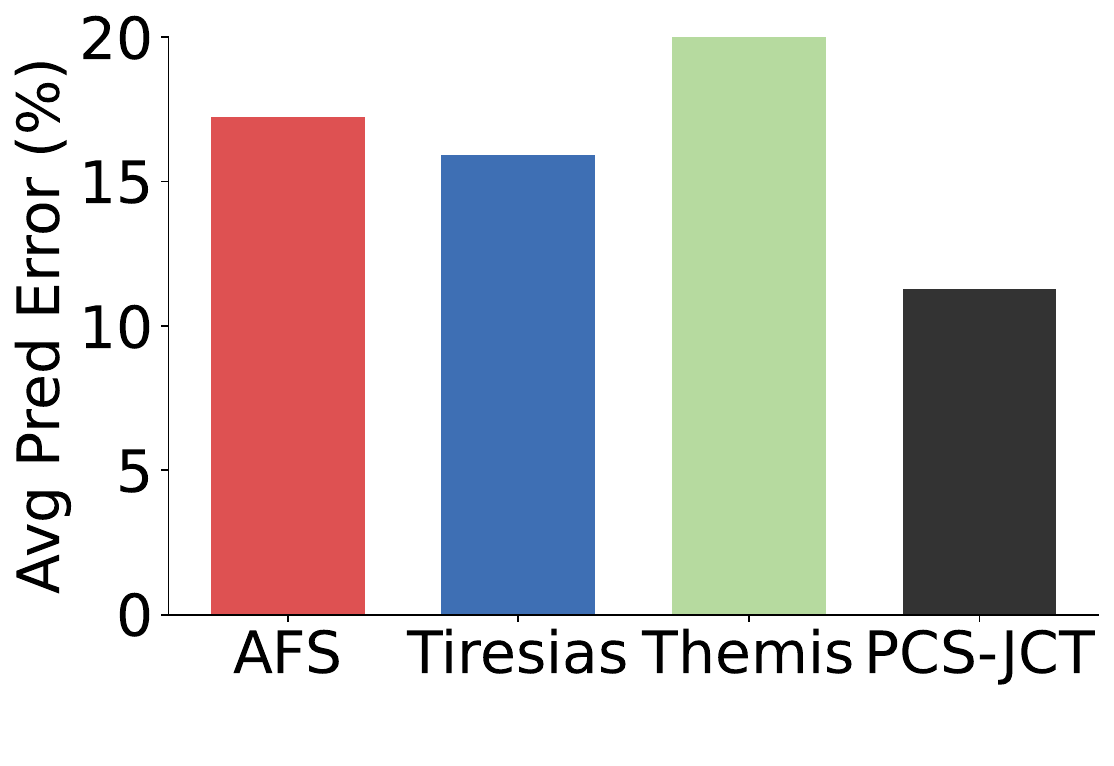}\label{subfig:sim-act-gavel}}
    \subfloat[][Pareto front]{\includegraphics[width=0.5\columnwidth]{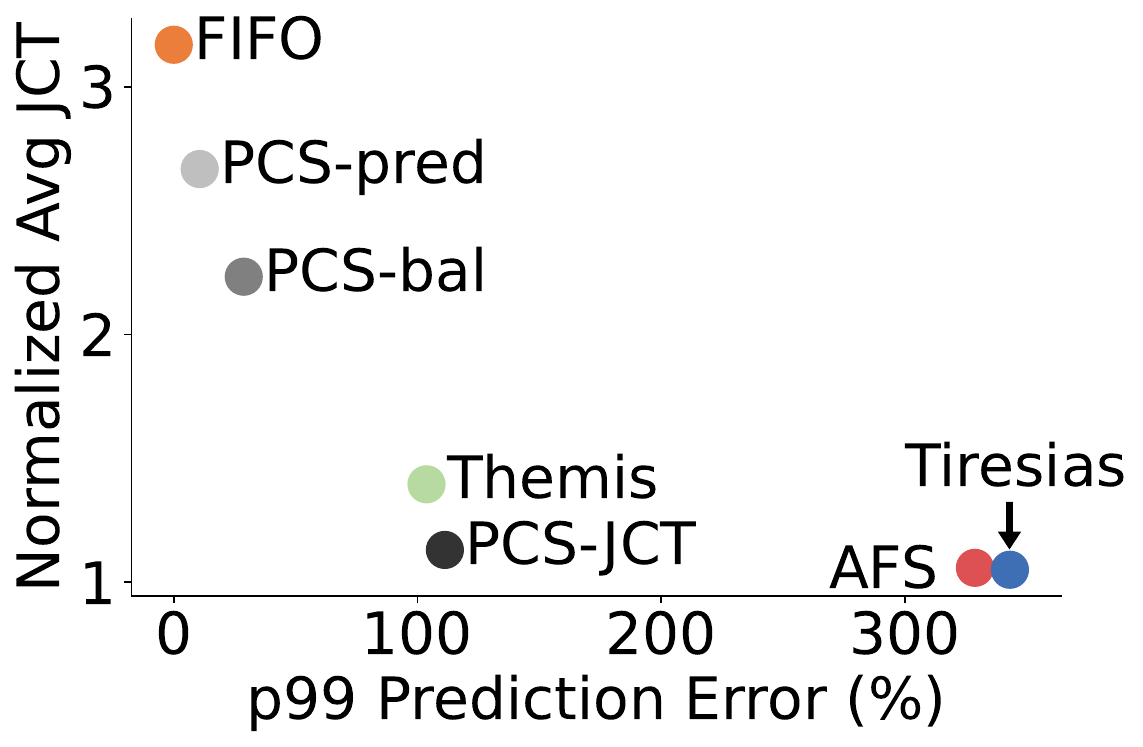}\label{subfig:sim-act-error-gavel}}
    \caption{[\textsc{Sim}]: Showing that schemes that optimize for average JCTs for workload-3 also have a small average error.
    For such workloads, the tail \ecte{} becomes an important metric.}
    \label{fig:sim-perf-vs-pred-gavel}
\end{figure}

\paragraph{Workload-2.}
Figure~\ref{fig:traces} compares the performance and predictability of \sys{} with other schedulers for workload-2.
For such workloads, AFS achieves the lowest possible avg JCT by giving more GPUs to jobs with higher throughput.
Despite its conservative approach to deal with sub-linear scaling jobs, \sys{} remains within 1.5 to 4$\ttimes{}$ of the optimal scheme for minimizing avg JCT, while drastically reducing the avg and tail \ecte{}.
For example, \sys{}-JCT reduces the average \ecte{} from 80\% to 1\% for trace \textsf{b436b2} and \sys{}-pred reduces the p99 \ecte{} from $900\%$ to $10\%$ for trace \textsf{6214e9}.

\paragraph{Workload-3.}
Figure~\ref{fig:sim-perf-vs-pred-gavel} compares the different schedulers for workload-3.
For this workload, we observe that schedulers optimized for performance, including \sys{}-JCT achieve reasonably low average \ecte{}.
This is because for workload-3, majority of the jobs are small and similar in size.
For such workloads, tail \ecte{}, becomes important owing to some jobs being starved under priority schedulers.
With the appropriate preference specification, \sys{} discovers configurations that can drastically reduce the p99 \ecte{}.
For example, \sys{}-JCT reduces the \ecte{} from $\geq{}300\%$ to $\approx{}100\%$ while being within $1.1\times{}$ of performance optimal schemes~(Fig.~\ref{subfig:sim-act-error-gavel}).

\subsection{Micro benchmarks}
\label{subsec:micro}

\begin{figure}[!t]
    \centering
    \subfloat[][Single sim time]{\includegraphics[width=0.32\columnwidth]{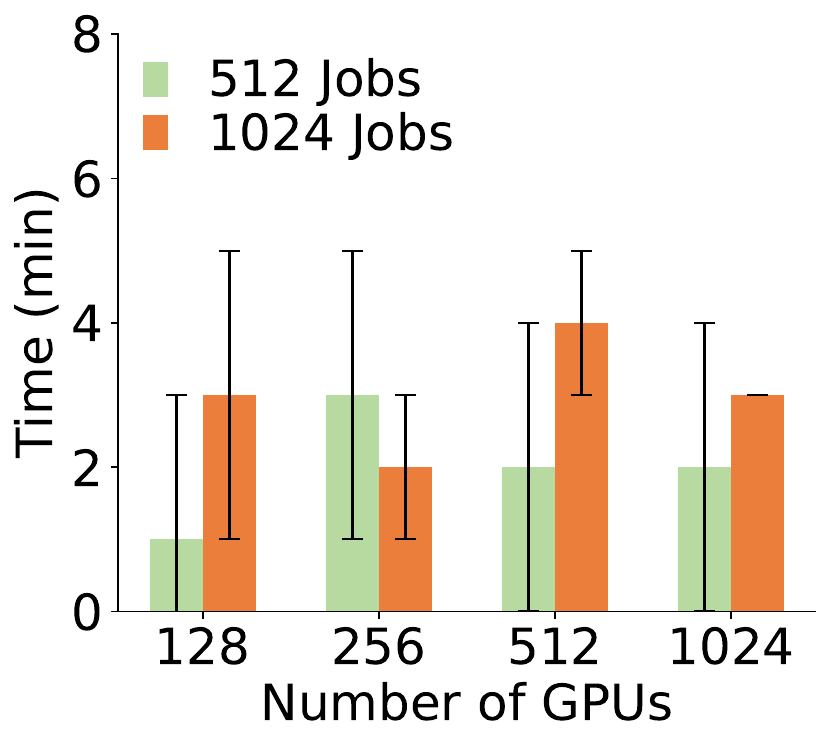}
    \label{fig:micro-single-time}}
    \subfloat[][Search time]{\includegraphics[width=0.32\columnwidth]{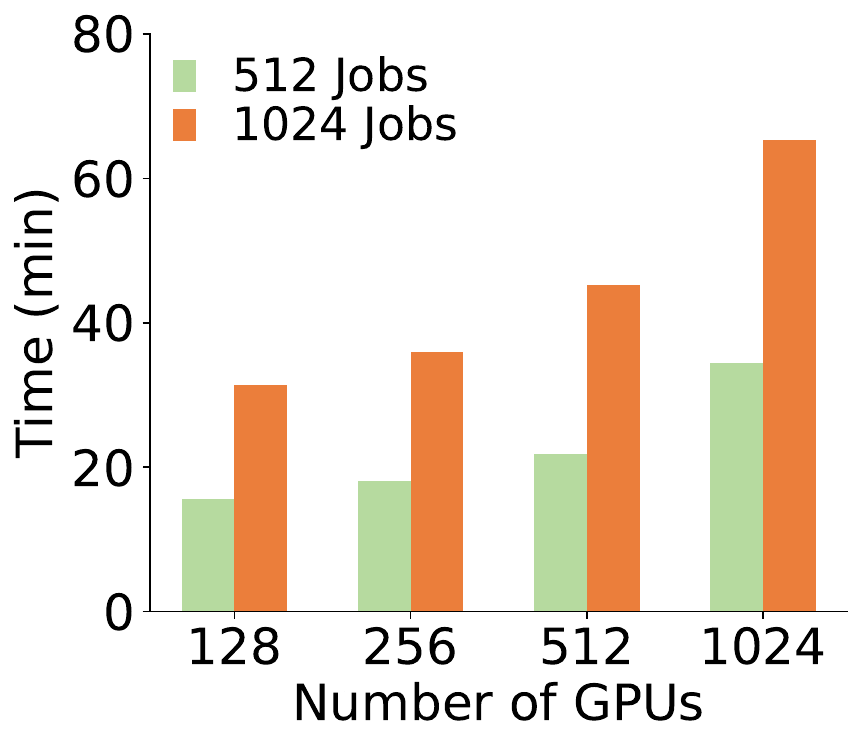}
    \label{fig:micro-pareto-time}}
    \subfloat[][Heuristics]{\includegraphics[width=0.32\columnwidth]{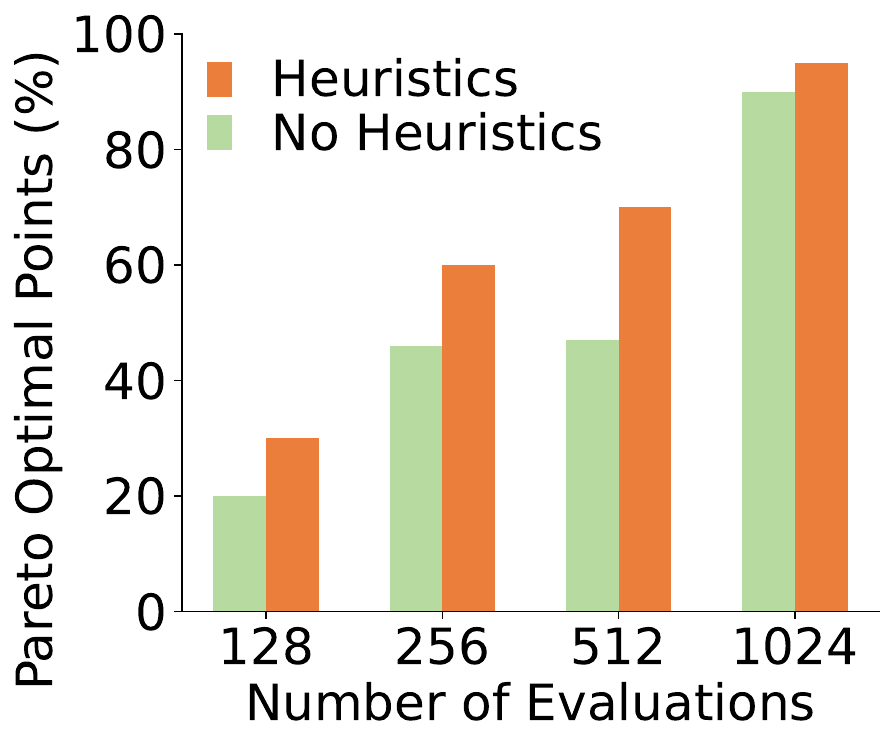}
    \label{fig:micro-sim-heuristics}}

    \caption{Feasibility of the simulation-based search strategy.
    (a) captures the time to run a single simulation, (b) shows the time it takes to discover the entire Pareto-front. (c) highlights that intelligent parameterization helps in discovering more Pareto optimal points for a given evaluation budget.}
    \label{fig:sim-performance}
\end{figure}

\paragraph{Feasibility of the search strategy.}
Figure~\ref{fig:micro-single-time} shows \sys{} takes $O(minutes)$ to run a single simulation for a given load (number of jobs) and cluster size (number of GPUs).
\sys{} extensively leverages the underlying parallelism to discover the entire Pareto-front -- requires running $\approx 1000$ simulations-- in approximately 60 minutes (Fig.~\ref{fig:micro-pareto-time}).
Figure~\ref{fig:micro-sim-heuristics} shows that \sys{} benefits from the heuristics (discussed in \S\ref{subsec:preference_solver}) to speed up the search and improve the quality of the Pareto-front by discovering new points that randomly sampling the search space takes more time to.


\begin{figure}[!t]
    \centering
    \subfloat[][\ecte{}]{\includegraphics[width=0.33\columnwidth]{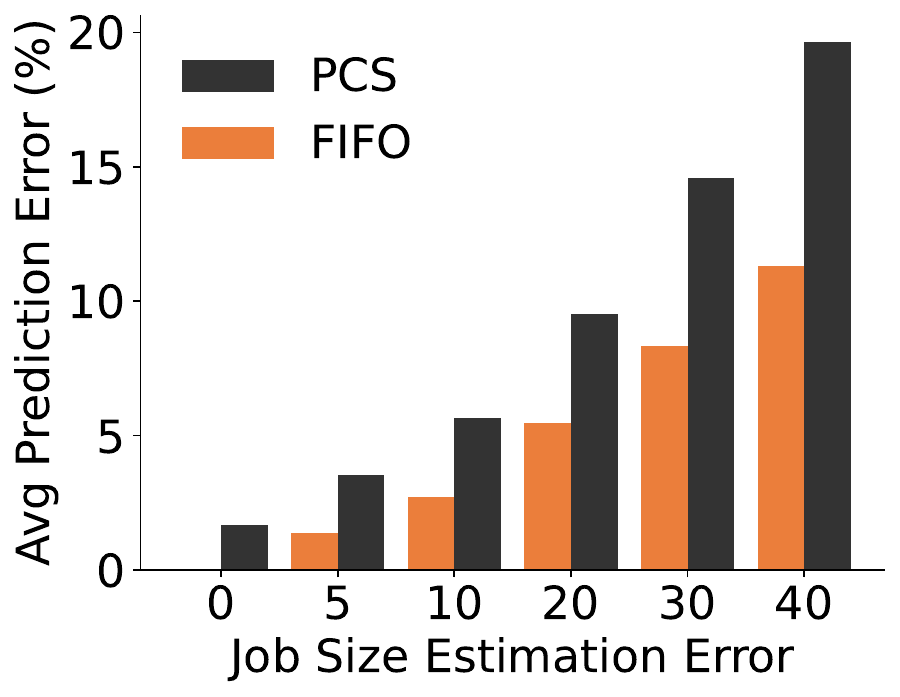}\label{fig:micro-error-pred}}
    \subfloat[][Avg JCT]{\includegraphics[width=0.33\columnwidth]{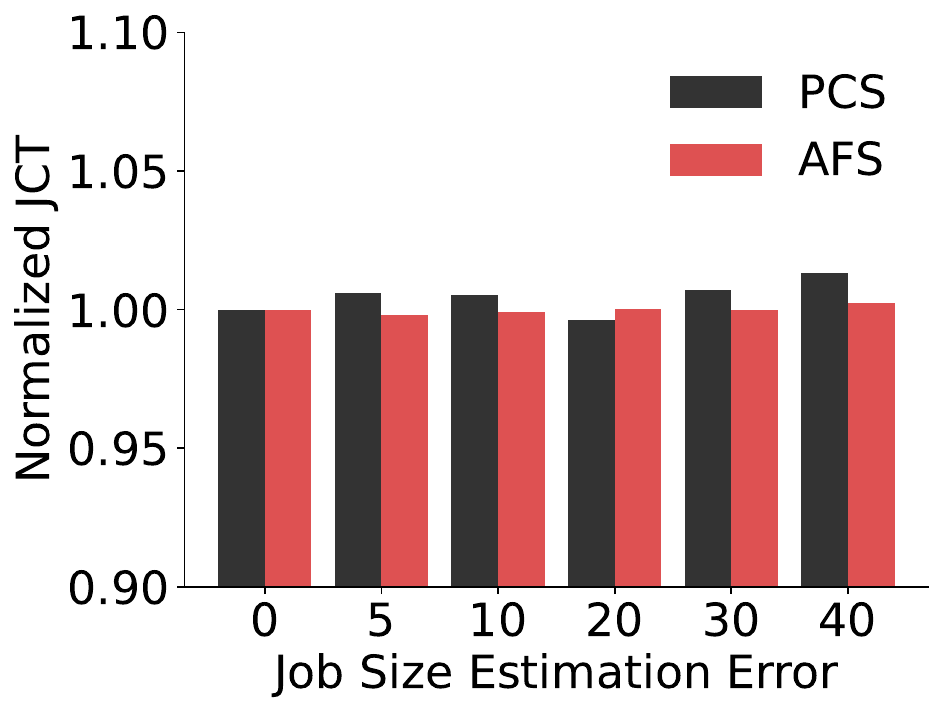}\label{fig:micro-error-JCT}}
    \subfloat[][Sensitivity]{\includegraphics[width=0.33\columnwidth]{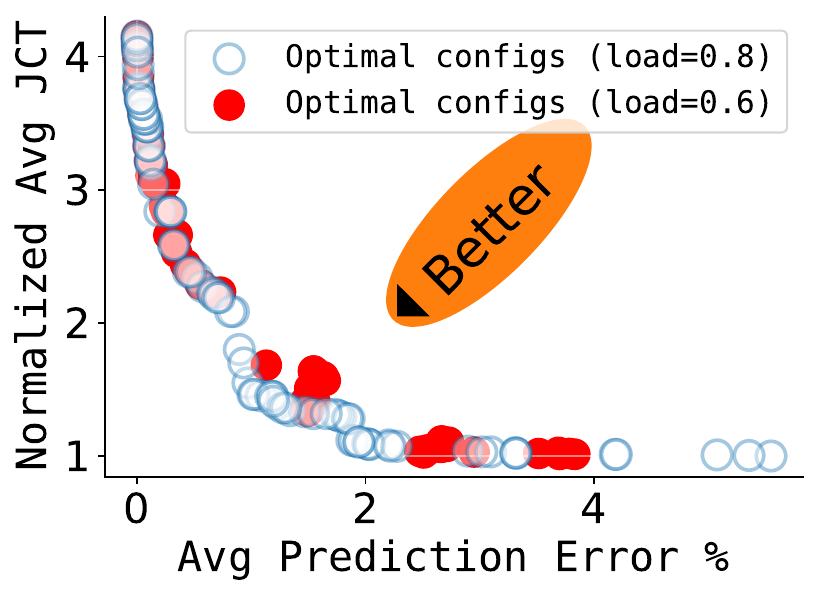}\label{fig:micro-error-load}}
    \caption{Shows the effects of error in job size and load estimation. a) compares the average \ecte{} using \sys{} and FIFO~\cite{YARN} with varying job size estimation error. b) compares the avg JCT of \sys{} and AFS~\cite{AFS} under the same error. c) shows sensitivity of WFQ configs to load changes.}
    \label{fig:micro-error}
\end{figure}

\paragraph{Error in job size estimation.}
Figure~\ref{fig:micro-error} shows the impact of estimation error in job-sizes on the predictability and performance of \sys{}.
As job-size estimation gets poorer, the impact on avg \ecte{} follows the same trend as the \ecte{} under FIFO (Fig.~\ref{fig:micro-error-pred}).
Figure~\ref{fig:micro-error-JCT}, compares the avg JCTs of AFS with no error in job-size estimation to \sys{} with varying estimation error.
\sys{} is still within $1.05\times{}$ of AFS.
This is because as long as the job is mapped to the correct class, the error in estimating its size has limited impact on performance.


\paragraph{Sensitivity of Pareto-optimal configurations.}
To evaluate the sensitivity of Pareto-optimal configurations, we evaluate configurations discovered for workload-1 assuming 60\% load on a system actually running at 80\% load.
Figure~\ref{fig:micro-error-load} shows that while the exact trends do not hold when the estimated workload is a mismatch, $75\%$ of the configurations are within $10\%$ of the closest Pareto-optimal point.

\section{Discussion}\label{sec:discussion}

\begin{figure}[!t]
    \centering
    \subfloat[width=0.5\columnwidth][Job size vs \ecte{}]{\includegraphics[width=0.495\columnwidth]{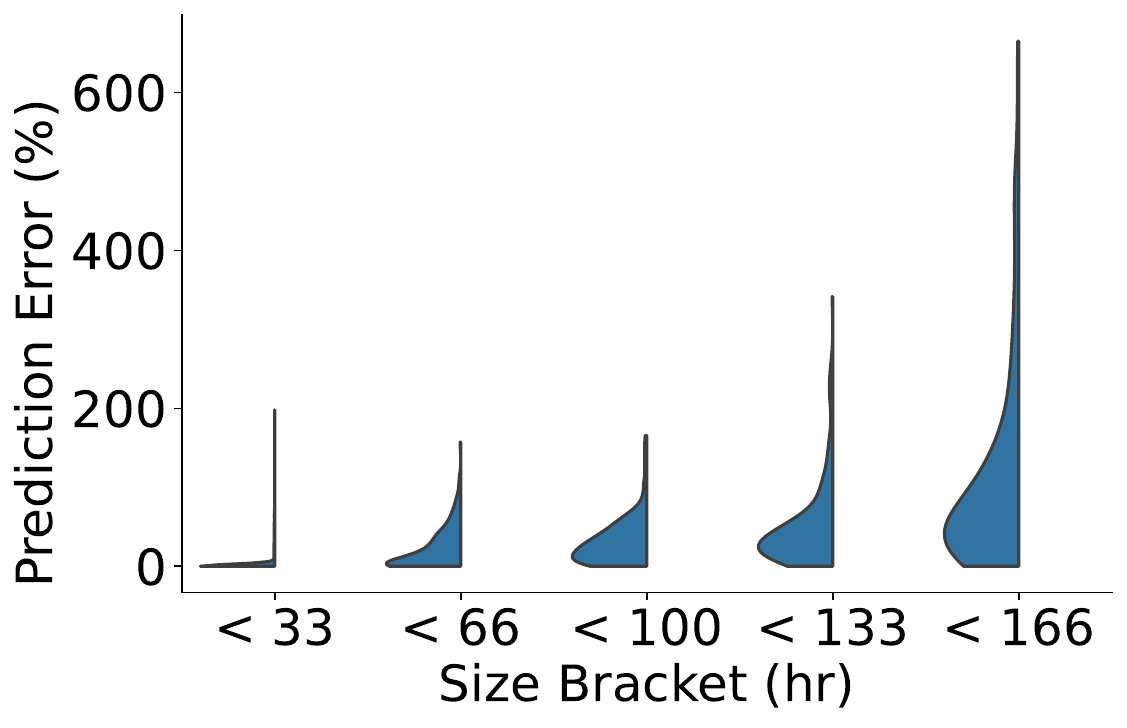}\label{fig:size_vs_estimate}}
    \subfloat[width=0.5\columnwidth][Contention observed by jobs]{\includegraphics[width=0.495\columnwidth]{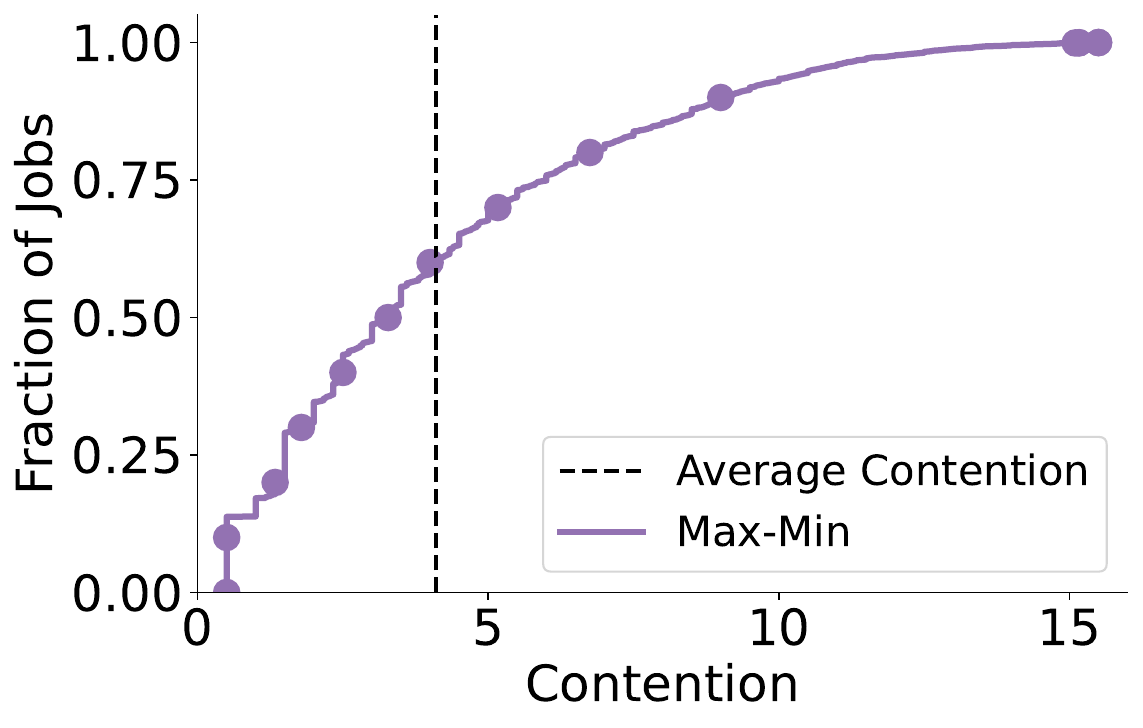}\label{fig:job_id_vs_apps_seen}}
    \caption{Limitation of size and contention based heuristics to predict JCTs (workload-3).
    For Tiresias (a), the correlation between job size and \ecte{} is weak as jobs of all sizes can suffer from unpredictability.
    For Max-Min (b), the contention observed by jobs shows variability, ruling out contention based heuristics like fair-finish-time to provide \cte{}.}
    \label{fig:correlation}
\end{figure}

\paragraph{Sophisticated prediction techniques.}
Using more complicated prediction techniques is orthogonal work.
We posit that future arrivals may be difficult to take into account in the prediction decision given that various attributes about them are unknown.
It is unclear how well sophisticated prediction techniques will generalize to different scenarios.
For instance, a future job's demand function cannot be determined before it actually arrives.
Our emphasis is on making scheduling \emph{predictable} and rely on a simple prediction strategy instead.

To highlight this point, we test the suitability of two simple scheduler specific heuristics --- job size and observed contention --- as features for a predictor.
For example, for Tiresias~\cite{tiresias}, which prioritizes small jobs, a predictor could leverage the job's size.
Similarly, the average contention faced by jobs can be used for fairness based policies.
Figure~\ref{fig:size_vs_estimate} shows that even small jobs under Tiresias can have \ecte{} up to $200\%$, whereas for long jobs it goes above $600\%$.
Similarly, figure~\ref{fig:job_id_vs_apps_seen} shows that the average observed contention is ~$4\times$ less than the contention observed at the tail under Max-Min scheduling.

\paragraph{Deciding between Pareto-optimal choices.}
Exposing trade-offs as Pareto-optimal choices can help operators to make informed choices by narrowing down the possibilities.
We still, however, rely on the operator's ability to decide between them.
Extending this to allow individual users to pick different preferences on a per job basis, while powerful, can result in cross-user conflicts which may be difficult to resolve.
We leave picking preferences on a per-job basis as future work.

\section{Related work}
\label{sec:related-work}

\paragraph{Scheduling systems.}
A large body of work emphasizes on intelligent GPU scheduling for DNN workloads, considering metrics such as minimizing average job completion times~\cite{salus,sidecar, allox,tiresias, AFS}, maximizing fairness~\cite{gandiva, gandiva_fair, themis}, cluster efficiency~\cite{AFS,allox, antman} and average DNN accuracy~\cite{slaq, optimus}.
They use preemption based techniques to achieve their objectives, which, we show in this paper, is detrimental to predictability.

\sys{} can benefit from system-level techniques such as elastic scaling~\cite{AFS,resource-elastic-mlsys}, GPU preemption~\cite{gandiva,wu2023transparent,thorpe2022bamboo,antman}, DNN throughput profiling~\cite{pollux,liu2021bgl}, job/AutoML app size estimation~\cite{themis, optimus}, and sharing-safety~\cite{hiveD} used in these systems.
However, in contrast to them, the scheduling policy, \policy{}, focuses on predictability by limiting the use of preemption and the flexibility to cluster operators in choosing various trade-off points between predictability and other traditional objectives.
Gavel~\cite{gavel} also translates different scheduling policies to optimization objectives but does not cover predictability and only finds a point solution for each objective while \sys{} allows operators to choose from a range of Pareto-optimal choices.

Deadline based schedulers~\cite{chronus, hypersched, seer} prioritize training jobs based on user provided deadlines.
This requires users to come up with reasonable deadlines.
If every user acts greedily and picks a tight deadline, the system can falter.
Secondly, aggressively meeting deadlines or discarding jobs whose deadlines cannot be met, fail to provide any utility/information to the user.
Instead of relying on deadlines and classifying them as soft/hard, we directly provide a \cte{} to the user.
%

\paragraph{Multi-class scheduling.}
A broad category of schedulers use the idea of class-based scheduling~\cite{2D, baraat, pias, aalo, lps, tiresias, pase, preemptionglobecom} in different contexts to achieve performance related goals.
We borrow ideas from these techniques.
For example, like 2D~\cite{2D} we also create classes based job size variation within a class.
However, these techniques opt for a fixed strategy in creating classes, mapping jobs to classes and assigning class weights (e.g., Baraat and Tiresias only use 2 classes) and will be limited to offering a fixed trade-off between objectives.

\paragraph{Adaptive schedulers.}
There are multiple recent examples of empirical, adaptive cluster management.
For example, SelfTune \cite{selftune} applies reinforcement learning techniques to automatically update the cluster management policy based on periodic cluster status updates.
Decima~\cite{sim-learning-scheduling-algorithms} uses simulations to learn optimal scheduling algorithms for data processing.
SWP~\cite{sim-swp} uses a simulation guided approach to find optimal bandwidth scheduling decisions.
Our strategy is inspired by these works.



\paragraph{Predictable scheduling.}
Predictable scheduling and delay guarantees has been studied in broader contexts.
Weirman et al~\cite{predictability-of-sched-policies} classify different scheduling policies based on the variation in the slowdown experienced by jobs.
Other studies~\cite{delay-information, how-tolerable-is-delay} look at the benefits of providing delay information to users and understand how much delay is tolerable.
CFQ~\cite{CFQ} defines predictability as a job's fair-finish-times (FFT), similar to Themis.
However, FFT is prone to variation itself as new jobs arrive~\cite{morpheus}.


\section{Conclusion}
In this paper, we call for providing predictability as a first order consideration in cloud scheduling systems.
Our inspiration comes from real-world systems that provide their users with predictions (e.g., estimated delivery dates).
Our solution, \sys{}, provides predictability while balancing other considerations like performance and fairness.
It comprises of a bi-directional preference interface to empower cloud operators in making informed trade-offs between multiple objectives.
To realize these trade-offs, \sys{} uses WFQ in unique way coupled with a simulation-based strategy to discover Pareto-optimal WFQ configurations.
Our results show the flexibility of \sys{} in achieving a wide range of operator objectives, offering a first step towards predictable scheduling in a practical way.
\emph{This work does not raise any ethical considerations.}
\newpage
\section*{Acknowledgements}
We thank all the anonymous reviewers for their feedback and all the member of the NAT lab for their valuable insights in helping us improve this work.
This work was supported by NSF CNS under award numbers 1815046 and 2106797.

\bibliographystyle{plain}
\bibliography{cleaned_ref}
\footnotesize{}



\end{document}